\documentclass[hidelinks,a4paper,10pt,twocolumn, switch]{article}

\usepackage{fancyhdr}
\fancypagestyle{plain}{%
  \fancyhf{}
\fancyfoot[R]{\thepage}
\fancyfoot[L]{\small \textit{Preprint submitted to Springer}\\ \texttt{arXiv}}
}

\usepackage[affil-it]{authblk}
\makeatletter
\def\@maketitle{%
  \newpage
  \null
  \vskip 2em%
  \begin{center}%
  \let \footnote \thanks
    {\Large\bfseries \@title \par}%
    \vskip 1.5em%
    {\normalsize
      \lineskip .5em%
      \begin{tabular}[t]{c}%
        \@author
      \end{tabular}\par}%
    \vskip 1em%
    {\normalsize \@date}%
  \end{center}%
  \par
  \vskip 1.5em}
\makeatother

\usepackage[a4paper, total={17.5cm,24cm}]{geometry}

\usepackage[font=normal,labelfont=bf]{caption}

\usepackage{sectsty} 

\usepackage{titlesec}
\titlespacing\section{0pt}{12pt plus 3pt minus 3pt}{1pt plus 1pt minus 1pt}
\titlespacing\subsection{0pt}{10pt plus 3pt minus 3pt}{1pt plus 1pt minus 1pt}
\titlespacing\subsubsection{0pt}{8pt plus 3pt minus 3pt}{1pt plus 1pt minus 1pt}

\titleformat{\section}{\normalfont\large\bfseries}{\thesection}{1em}{}

\titleformat{\subsection}{\normalfont\normalsize\bfseries}{\thesubsection}{1em}{}

\titleformat{\subsubsection}{\normalfont\normalsize}{\thesubsubsection}{1em}{}

\titleformat{\paragraph}[runin]{\normalfont\normalsize\itshape}{\theparagraph}{1em}{}

\usepackage{mathtools,upgreek}

\usepackage[utf8]{inputenc}	
\usepackage[T1]{fontenc}	
\usepackage{xcolor}		
\usepackage{booktabs} 		
\usepackage{nicefrac}		
\usepackage{microtype}		
\usepackage{lineno}		
\usepackage{float}			

\usepackage{nicefrac}
\usepackage{paralist}
\usepackage{graphicx}
\usepackage{caption}
\usepackage{placeins}
\usepackage{dblfloatfix}
\usepackage{subfigure}
\usepackage{amsmath}  
\usepackage{amsfonts} 
\usepackage{pifont} 
\usepackage{lineno}
\usepackage[linktoc=page,
colorlinks=true,	
linkcolor=blue,
citecolor=blue,
breaklinks=true]{hyperref}  

\usepackage{enumitem} 
\usepackage{stmaryrd}  

\usepackage{algorithm}
\usepackage{algpseudocode}
\usepackage{hyperref}

\usepackage{siunitx}
\usepackage{breakcites}

\usepackage{mathptmx}      
\usepackage{newtxtext,newtxmath}      
\usepackage{tabularx}
\usepackage{listings}
\mathtoolsset{centercolon}

\title{Reconstructing microstructures from statistical descriptors using neural cellular automata}

\begin{document}

\author{Paul Seibert${}^\mathrm{a}$, \quad Alexander Ra{\ss}loff${}^\mathrm{a}$, \quad Yichi Zhang${}^\mathrm{a}$, \quad Karl Kalina${}^\mathrm{a}$, \quad Paul Reck${}^\mathrm{b}$, \quad Daniel Peterseim${}^\mathrm{b}$, \quad Markus K\"astner${}^\mathrm{a, c}$
  \thanks{Contact: \texttt{markus.kaestner@tu-dresden.de} }}
\affil{${}^\mathrm{a}$Institute of Solid Mechanics, TU Dresden, Germany \\ ${}^\mathrm{b}$Institute of Mathematics, University of Augsburg, Germany \\ ${}^\mathrm{c}$Dresden Center for Computational Materials Science, TU Dresden, Germany}

\date{}
\maketitle

\begin{abstract}
The problem of generating microstructures of complex materials in silico has been approached from various directions including simulation, Markov, deep learning and descriptor-based approaches.
This work presents a hybrid method that is inspired by all four categories and has interesting scalability properties.
A neural cellular automaton is trained to evolve microstructures based on local information.
Unlike most machine learning-based approaches, it does not directly require a data set of reference micrographs, but is trained from statistical microstructure descriptors that can stem from a single reference. 
This means that the training cost scales only with the complexity of the structure and associated descriptors.
Since the size of the reconstructed structures can be set during inference, even extremely large structures can be efficiently generated.
Similarly, the method is very efficient if many structures are to be reconstructed from the same descriptor for statistical evluations.
The method is formulated and discussed in detail by means of various numerical experiments, demonstrating its utility and scalability.

\vspace{5mm}
\noindent
\textbf{Keywords: } Microstructure~\textendash~Reconstruction~\textendash~Descriptor~\textendash~Neural cellular automata

\end{abstract}

\section{Introduction}
\label{sec:introduction}

The generation and analysis of random heterogeneous composite materials is a recently emerging research topic that aims at accelerating materials engineering by enabling digital workflows such as numerical simulation and inverse design~\cite{chen_data-centric_2022}.
Specifically, microstructure characterization and reconstruction allows to 
\textit{(i)} generate many microstructure realizations from a single example,
\textit{(ii)} explore hypothetic materials by interpolating between microstructures in a morphologically meaningful manner, and
\textit{(iii)} create 3D models from 2D observations.
A multitude of approaches has been developed in the last decades that is summarized in different review articles~\cite{bargmann_generation_2018,bostanabad_computational_2018,sahimi_reconstruction_2021}.
For the purpose of this work, the existing approaches can be broadly divided in four categories\footnote{Besides hybrid methods that fall into multiple categories, some exceptions like Wang tiles~\cite{doskar_microstructure-informed_2021} do not clearly fall into any of the categories.} - simulation, Markov random field, deep learning and descriptor-based approaches.
Naturally, some algorithms in the literature can be identified as hybrid methods that fall into two or more of these categories.
After discussing the main ideas of these categories and approaches in \autoref{sec:introexisting}, this work presents an algorithm that bridges all four categories and exhibits some very interesting properties as described in \autoref{sec:intronca}.

\subsection{Existing approaches for microstructure reconstruction}\label{sec:introexisting}
\paragraph{Simulation-based approaches}
Simulating the microstructure evolution might be the most direct way.
This requires to identify and to solve the physical (partial differential) equations~(PDEs) that govern the process.
An excellent overview is given in~\cite{bargmann_generation_2018}.
As an example, the Cahn-Hilliard equation describing phase separation~\cite{cahn_free_1958} has been studied extensively~\cite{wight_solving_2021,prakasha_two_2019,risthaus_solving_2022}.
Similarly, for granular structures, given a representative set of particles, realistic and dense packing can be achieved by simulating gravitational forces~\cite{zhao_three-dimensional_2022,winkler_granular_2014,vlassis_synthesizing_2023}.
As a final, more complex example, grain formation in polycristalline structures has been studied in depth. 
Simplified approaches reduce the description to vertices~\cite{kawasaki_vertex_1989} or grain boundaries~\cite{brakke_surface_1992}, whereas Monte Carlo methods~\cite{anderson_computer_1989} or cellular automata~\cite{janssens_introductory_2010,zhan_cellular_2008,ghumman_calibration_2023} are used to model the evolution of an entire 2D pixel field.
Recently, neural cellular automata have been applied to solidification microstructure modeling~\cite{tang_neural_2023}.
Approaches based on the phase field method are probably the most developed.
Thereby, the evolution of a diffuse indicator function is modeled by an additional differential equation~\cite{boettinger_phase-field_2002,tourret_phase-field_2022,skogvoll_hydrodynamic_2022} that can be solved, for example, in \emph{OpenPhase}~\cite{tegeler_parallel_2017}.
These approaches are often applied to simulate the complex microstructure morphologies that arise in additive manufacturing~\cite{tan_microstructure_2020,korner_modeling_2020,li_solidification_2020}.
This non-exhaustive list indicates that a variety of physical processes are responsible for the formation of different material classes.
Even if the relevant set of physical equations is selected, it can be challenging to perform the simulations due to numerical issues or difficulties in parameterizing the underlying constitutive models~\cite{yan_data-driven_2018,korner_modeling_2020}.
This motivates the purely image-based approaches that are presented in the following.

\paragraph{Markov-based reconstruction}
As a first purely image-based method, this subsection discusses a class of reconstruction algorithms originally developed for computer graphics applications which are herein referred to as Markov-based approaches.
For this purpose, it is worth noting that a microstructure can be modeled as a stationary Markov random field if the probability of finding a certain phase at a given location does not depend directly on the location, but only on the phase distribution in the local finite-size neighborhood.

This assumption of locality and stationarity motivates reconstruction algorithms that directly rely on this conditional probability to update individual pixels based on their neighbor's values. 
A very simple implementation inspired by texture synthesis~\cite{wei_state_2009} might determine individual pixel updates by scanning the reference data for the given neighborhood in order to compute a probability~\cite{sundararaghavan_reconstruction_2014,liu_random_2015}.
It is worth noting that this approach is akin to the multi-point statistics method that has been developed in the Geosciences literature~\cite{strebelle_conditional_2002} and has been applied and improved substantially by Tahmasebi~\cite{tahmasebi_cross-correlation_2013,tahmasebi_enhancing_2016,tahmasebi_enhancing_2016-1,tahmasebi_stochastic_2018}.
For a better scalability, improved algorithms precompute the probabilities for all neighborhoods and store them in efficient data structures for access during reconstruction~\cite{strebelle_conditional_2002,straubhaar_improved_2011}.
Direct sampling methods~\cite{mariethoz_direct_2010} as well as data structure-based alternatives are implemented in \emph{MPSLIB}~\cite{hansen_mpslib_2016}.

Despite a good local prediction quality, MRF-based approaches often fail to accurately reproduce long-range correlations.
This behavior is related to the neighborhood size in the Markovian assumption: 
Capturing long-range correlations requires large neighborhood sizes, which are often unfeasible because of a disproportionately increased need for training data.
Multigrid approaches~\cite{fu_hierarchical_2023,tahmasebi_stochastic_2018} have been shown to alleviate this issue to a certain extent.
Furthermore, to condense the information to a compact model that is also able to interpolate missing neighborhood patterns from similar examples, supervised models have been trained to predict a pixel's phase given its neighborhood.
In particular, decision trees~\cite{bostanabad_stochastic_2016} and neural networks~\cite{latka_microstructure_2021,noguchi_stochastic_2021,fu_hierarchical_2023} have been used for 2D and 3D~\cite{bostanabad_characterization_2016} reconstruction.
This motivates the discussion of purely deep learning-based approaches in the following subsection.

\paragraph{Deep learning-based reconstruction}
In deep learning-based methods, a generative model is fitted or trained on a sufficiently large data set of microstructures and is then used to sample new realizations of the same structure.
Autoencoders~\cite{cang_microstructure_2017,noguchi_stochastic_2021,faraji_niri_performance_2022} and generative adversarial networks (GANs) are typical examples that have been applied to MCR~\cite{mosser_reconstruction_2017,li_deep_2018}.
For the latter, the merits of modifications like conditional GANs~\cite{iyer_conditional_2019,feng_reconstruction_2019}, SytleGAN~\cite{fokina_microstructure_2020}, and gradient penalty~\cite{li_digital_2022} have also been discussed in the context of microstructure generation. 
Applications to steels~\cite{lee_virtual_2021} and earth materials~\cite{amiri_quantifying_2022} show high image quality.
Although GANs usually operate on 2D data, 3D-to-3D reconstruction can be achieved by using 3D convolutions~\cite{hsu_microstructure_2020,henkes_three-dimensional_2022}.
For reconstructing 3D data from 2D examples, a 3D generator has been combined with a 2D discriminator~\cite{coiffier_3d_2020,kench_generating_2021}.
As an alternative, the third dimension can be regarded as time by combining the GAN with a recurrent neural network~\cite{zhang_pm-arnn_2023}.
To harness the advantage of both, autoencoders and GANs, they are sometimes combined by using the decoder simultaneously as a generator.
This has proven advantageous for 2D-to-3D reconstruction~\cite{shams_coupled_2020,feng_end--end_2020,zhang_slice--voxel_2021} and for extremely small data sets~\cite{zhang_da-vegan_2023}.

As an alternative, machine learning methods like Bayesian approaches~\cite{chamani_rapid_2023} and attention-based models~\cite{zhang_3d_2022,zheng_rockgpt_2022,phan_size-invariant_2022} are equally applicable.
Diffusion models, which have recently replaced GANs as state-of-the-art in general-purpose image generation, have also been applied to microstructure reconstruction~\cite{dureth_conditional_2023,lee_microstructure_2023} and optimization~\cite{lim_microstructure_2023,vlassis_denoising_2023}.

Much research is focused on identifying suitable model types and adapting them to microstructure reconstruction by enabling 2D-to-3D reconstruction~\cite{zhang_slice--voxel_2021,kench_generating_2021} making them applicable to small data sets~\cite{zhang_da-vegan_2023} or ensuring that certain descriptor requirements are met~\cite{li_transfer_2018,robertson_local-global_2023}.
A major challenge lies in defining models with high accuracy that at the same time do not require large data sets to be trained on. These challenges motivate training-free models such as descriptor-based reconstruction, as presented in the next subsection.

\paragraph{Descriptor-based reconstruction}
The central idea behind descriptor-based reconstruction methods is to statistically quantify the microstructure morphology by means of descriptors like volume fractions and spatial $n$-point correlations~\cite{torquato_random_2002}.
Reconstructing a microstructure from a given set of descriptors can then be formulated as an optimization problem directly in the space of possible microstructures.
Here, the desired microstructure descriptors can be computed from a single microstructure example, making these methods very data-efficient.

One of the most well-known descriptor-based reconstruction methods is the Yeong-Torquato algorithm~\cite{torquato_random_2002}, which iteratively swaps individual pixels in the microstructure to solve the optimization problem.
A detailed discussion is given in~\cite{jiao_modeling_2007,jiao_modeling_2008}.
This enables high flexibility, as descriptors can be replaced by new alternatives~\cite{feng_reconstruction_2018,piasecki_statistical_2018} or higher-fidelity versions of the same descriptor~\cite{gerke_improving_2014,seibert_relevance_2023}.
However, even with computationally inexpensive descriptors, the Yeong-Torquato algorithm becomes computationally challenging at high resolutions and in 3D, where billions of iterations are sometimes required for convergence~\cite{adam_efficient_2022}.
A common solution is to use a multigrid scheme~\cite{alexander_hierarchical_2009,pant_multigrid_2015,karsanina_hierarchical_2018,chen_fast_2022,seibert_two-stage_2023}.
Further ideas include different-phase neighbor sampling rules~\cite{pant_stochastic_2014}, efficient descriptor updates~\cite{rozman_efficient_2001,adam_efficient_2022} and optimized directional weighing of correlation functions~\cite{gerke_improving_2014}.
More information is given in~\cite{bostanabad_computational_2018}.

As an alternative to the pixel-based Yeong-Torquato algorithm, the optimization problem can be formulated in a much lower-dimensional space.
For this purpose, the microstructure is approximated by geometric objects that can be described by a few parameters, e.g., ellipsoidal inclusions~\cite{xu_descriptor-based_2014,scheunemann_design_2015,seibert_fast_2023} or Voronoi or Laguerre cells~\cite{groeber_framework_2008,quey_neperfepx_2022,prasad_kanapy_2019}.

Independently from the microstructure representation~\cite{seibert_fast_2023}, {differentiable} descriptors allow solving the optimization problem using a gradient-based optimizer.
This idea is formulated as differentiable microstructure characterization and reconstruction (DMCR)~\cite{seibert_reconstructing_2021,seibert_descriptor-based_2022} and several approaches can be identified as special cases~\cite{li_transfer_2018,bostanabad_reconstruction_2020,bhaduri_efficient_2021}.

The Yeong-Torquato algorithm and improved versions of it, such as DMCR, have been successfully validated and applied to alloys and anisotropic metamaterials~\cite{seibert_two-stage_2023} sandstone~\cite{zhou_3d_2018}, rock~\cite{xiao_fracture_2021}, chalk~\cite{talukdar_stochastic_2002}, various soils~\cite{gerke_description_2012} and more.
Some versions are publicly available in the open-source \emph{MCRpy} package~\cite{seibert_microstructure_2022}.

While descriptor-based approaches are very accurate and data-efficient since no training data set is required, they are computationally intensive.
More specifically, since the optimization is directly carried out in the microstructure space, the memory and computational requirements grow quickly as the microstructure size increases, especially in 3D.

\paragraph{Hybrid reconstruction approaches}
The specific and unique advantages and disadvantages of all four categories of MCR approaches motivate hybrid methods that fall into multiple of these categories.
Naturally, there is no sharp boundary between Markov-based and deep learning methods if a machine learning model like a neural network is used to predict individual pixels based on their neighborhood as in~\cite{bostanabad_stochastic_2016,latka_microstructure_2021,noguchi_stochastic_2021,fu_hierarchical_2023,bostanabad_characterization_2016}.
Furthermore, simulation by discretized (partial) differential equations and cellular automata resemble Markov-based methods in their locality, but are derived from physical principles and sometimes incorporate various physical quantities (e.g. temperature) beyond phase indicator functions.
At the boundary between machine learning and descriptor-based methods, multiple sequential approaches use Gaussian random field-based methods~\cite{robertson_efficient_2021} to initialize simulated annealing\footnote{This is technically a hybrid method between two descriptor-based approaches.}~\cite{talukdar_stochastic_2002,jiang_efficient_2013} and diffusion models~\cite{robertson_local-global_2023}.
Furthermore, the volume fractions~\cite{zhang_fast_2022,robertson_local-global_2023,zhang_pm-arnn_2023,su_microstructure_2022,feng_reconstruction_2019}, histograms~\cite{li_cascaded_2022} and Gram matrices~\cite{zhao_three-dimensional_2022,yang_microstructural_2018} are sometimes added to the loss function of deep learning-based methods as microstructure descriptors.
\emph{DRAGen}~\cite{henrich_dragen_2023} combines an automaton-like growth process with a nucleation point optimization based on classical descriptors and allows to use machine learning models for generating input data.
At the interface between machine learning and physical simulation, autoencoders~\cite{macedo_what_2023} and diffusion models~\cite{vlassis_synthesizing_2023} have been used as particle generators followed by a gravity simulation for aggregate structures.
Besides that, the literature comprises a large number of physics-informed neural network approaches that are not discussed herein.

\subsection{Objectives and contribution of this work}\label{sec:intronca}
This work presents a hybrid approach that is inspired by all these categories.
Like in a simulation-based approach, a partial differential equation models the temporal evolution of the microstructure.
It is, however, not derived from physics but learned by a neural network.
Similar to the Markov-based methods, this network operates based on local information and is therefore called neural cellular automaton (NCA).
This constraint of locality is relaxed not by increasing the neighborhood beyond a one pixel distance, but by introducing further hidden channels to the microstructure function that the NCA can use to encode relevant information.
Finally, unlike common machine learning or Markov-based approaches, the NCA is not trained directly on image data or on a set of neighborhoods, but on a statistical descriptor.
This requires the NCA to be retrained whenever the statistical descriptor changes, however, it reduces the amount of required data to a bare minimum.
The input image only needs to enable the computation of a statistical descriptor; hence the NCA is applicable whenever classical training-free approaches like the Yeong-Torquato algorithm and DMCR can be used.
Furthermore, the size of the training data is independent of the image size during training, which is again independent from the size of the reconstructed structure.
Hence, microstructures of massive resolutions or numbers can be reconstructed with very limited additional computational effort.
Furthermore, due to the nature of NCA, the algorithm is inherently distributed, parallel and robust with respect to perturbations.

In summary, the central idea lies in modeling the differential equation governing the structure evolution by training neural cellular automata (NCA) on statistical descriptors.
A detailed formulation is given in \autoref{sec:nca} and validated by various numerical experiments in \autoref{sec:numexp}.
A conclusion is drawn in \autoref{sec:conclusion}.

\section{Neural cellular automata for descriptor-based microstructure reconstruction}\label{sec:nca}
Based on the work of Mordvintsev et al.~\cite{mordvintsev_growing_2020}, the formulation of general neural cellular automata (NCA) is summarized in~\autoref{sec:ncaformulation}.
The main idea of the present work to train NCA by arbitrary descriptors is described in \autoref{sec:ncadescriptor}.
Finally, the implementation is discussed in \autoref{sec:ncaimplementation}.

\subsection{Formulation of neural cellular automata}\label{sec:ncaformulation}
The general idea behind a cellular automaton is to iteratively update individual pixels based on the direct neighbors.
In the work of Mordvintsev et al.~\cite{mordvintsev_growing_2020}, this information source is further restricted.
The neighboring pixel values are not passed directly to the cellular automaton.
Instead, they are used to compute a discrete approximation to the gradient and curvature, which are then passed to the cellular automaton.
Denoting~$\boldsymbol{x}\in\mathcal D$ as a position vector in the microstructure domain $\mathcal D\subset \mathbb R^2$ and~$t\in\mathcal T = \{t\in\mathbb R \, | \, 0 \le t \le t^\text{end}\}$ as time, the evolution of the microstructure $m(\boldsymbol{x}, t)$ can be written as a partial differential equation
\begin{equation}
    \dfrac{\partial {m}(\boldsymbol{x}, t)}{ \partial t} = f_{\boldsymbol{\theta}} \left(  {m}(\boldsymbol{x}, t), \nabla_{\boldsymbol{x}} {m}(\boldsymbol{x}, t), \nabla_{\boldsymbol{x}}^2  {m}(\boldsymbol{x}, t) \right) \; ,
    \label{eqn:pdenonhidden}
\end{equation}
where $f_{\boldsymbol{\theta}}$ is the cellular automaton which maps the value, gradient and curvature of the microstructure function to its temporal derivative.
To be more specific, $\nabla_x(\bullet)$ and~$\nabla_x^2(\bullet)$ denote the gradient and Laplace operator, respectively.
Furthermore,~$m$ takes real values within the arbitrarily chosen bounds
\begin{equation}
    0 \leq m(\boldsymbol{x}, t) \leq 1 \quad \forall \; \boldsymbol{x} \in \mathcal D, \; t\in\mathcal T \; .
    \label{eqn:bounds}
\end{equation}
In a neural cellular automaton specifically, a neural network is chosen as $f_{\boldsymbol{\theta}}$, where~$\boldsymbol{\theta}$ denotes the parameter vector.

In other words, the NCA defines partial differential equation (PDE)\footnote{To be precise, the NCA defines a PDE \emph{system}, as explained later in the document.} that needs to be discretized and solved in order to generate a microstructure.
An explicit Euler scheme is chosen as a time stepping scheme
\begin{equation}
    \dfrac{{m}_{n_t+1} - {m}_{n_t}}{ \Delta t} = f_{\boldsymbol{\theta}} \left(  {m}_{n_t}, \nabla_{\boldsymbol{x}} {m}_{n_t}, \nabla_{\boldsymbol{x}}^2  {m}_{n_t} \right) \; ,
    \label{eqn:pdenonhiddendiscretized}
\end{equation}
where the current solution at time step~$n_t$ defines the update for the next time step~$n_t+1$.
The dependence on~$\boldsymbol{x}$ and~$t$ is dropped for the sake of brevity.
The space is naturally discretized on an equidistant grid of pixel values, where $\nabla_x(\bullet)$ and~$\nabla_x^2(\bullet)$ are approximated by a Sobel and Laplace filter, respectively.
Based on this discretization, the relation between the current solution, its spatial derivatives and its temporal evolution, i.e. the PDE itself, is learned by the NCA.

Given the inability of Markov-based approaches with small neighborhood sizes to accurately capture long-range correlations, it should be clear that the extremely limited local information is not sufficient to train a good NCA.
For this reason, the augmented microstructure function~$\boldsymbol{m}'(\boldsymbol{x}, t)$ is introduced which maps a spatial position~$\boldsymbol{x}$ at time~$t$ to an~$n$-dimensional vector.
The first entry of the vector contains the normal microstructure function~$m(\boldsymbol{x}, t)$ and is the only entry that affects the training.
The idea behind the other entries is that the NCA can choose to allocate any quantity that is useful for passing information and increasing the image quality.
As an example, for an equidistant grain microstructure, one channel might contain the distance to the next grain boundary.
With this, the temporal evolution reads
\begin{equation}
    \dfrac{\partial \boldsymbol{m}'(\boldsymbol{x}, t)}{ \partial t} = f_{\boldsymbol{\theta}} \left(  \boldsymbol{m}'(\boldsymbol{x}, t), \nabla_{\boldsymbol{x}} \boldsymbol{m}'(\boldsymbol{x}, t), \nabla_{\boldsymbol{x}}^2  \boldsymbol{m}'(\boldsymbol{x}, t) \right)
    \label{eqn:pdefull}
\end{equation}

For reconstructing a microstructure from the trained NCA,~$\boldsymbol{m}'(\boldsymbol{x}, 0)$ is initialized by zeros and the system evolves freely.

\subsection{Training the model from microstructure descriptors}\label{sec:ncadescriptor}
\begin{figure*}[htpb]%
\centering
\includegraphics[width=0.9\textwidth]{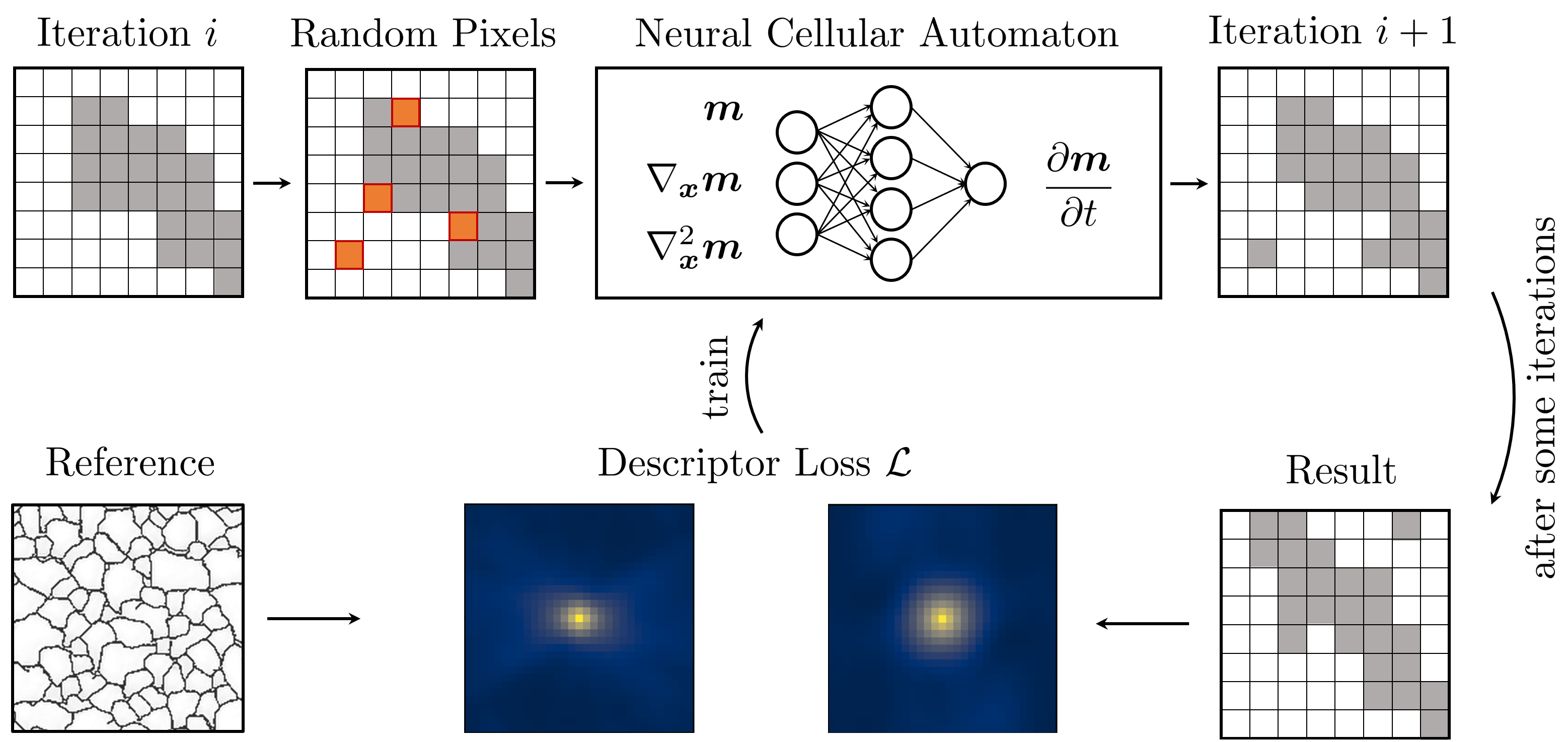}
\caption{Training procedure for a neural cellular automaton (NCA): In every iteration $i$, random pixel locations are chosen where the gradient and curvature is computed numerically. Together with the pixel value, these quantities are given to the NCA to predict a pixel update. After some time increments, the result is compared to the reference to train the NCA. This comparison is only carried out in terms of statistical descriptors.}\label{fig:graphical_abstract}
\end{figure*}

The function~$f_{\boldsymbol{\theta}}$ is learned by a small neural network with two layers as shown in \autoref{fig:graphical_abstract}:
\emph{First}, an initial solution $\boldsymbol{m}'(\boldsymbol{x}, 0) = \boldsymbol{0} \; \forall \; \boldsymbol{x}$ is chosen.
\emph{Secondly}, $\boldsymbol{m}'(\boldsymbol{x}, t)$ develops according to \autoref{eqn:pdefull} for a randomly chosen number of time steps.
As a regularization and as a measure to break symmetry, asynchronous updates are chosen, whereby in every time step, a given percentage of cells is chosen at random and only those develop.
The bounds given in \autoref{eqn:bounds} are enforced by clipping.
\emph{Thirdly}, a loss function~$\mathcal{L}$ is computed on the final result~$m^\mathrm{end} = m(\boldsymbol{x}, t^\mathrm{end})$.
The choice of~$\mathcal{L}$ is discussed later.
Note that only $m^\mathrm{end}$, i.e., the first component of $\boldsymbol{m}'{}^\mathrm{end}$, contributes to~$\mathcal{L}$.
\emph{Finally}, the gradient $\partial \mathcal{L} / \partial \boldsymbol{\theta}$ of the loss function with respect to the NCA parameters is computed by conventional backpropagation and used to update~$\boldsymbol{\theta}$.
Note that this limits the number of timesteps during training for numerical reasons.

The formulation of the loss function depends on the area of application of the NCA.
After initially using a pixel-wise Euclidean norm error in the RGB space for general-purpose image generation~\cite{mordvintsev_growing_2020}, Mordvintsev et al.~\cite{mordvintsev_texture_2021} found that a Gram matrix-based style loss~\cite{gatys_texture_2015} enable NCAs to be applied to texture synthesis~\cite{mordvintsev_texture_2021}.

The novelty in the present work lies in realizing that any of the known statistical descriptors can be used, as long as they can be differentiated with respect to the microstructure field.
The loss is thus formulated as a mean squared error~(MSE) in the descriptor space
\begin{equation}
    \mathcal{L} = \| \boldsymbol{D}(m^\mathrm{end}) - \boldsymbol{D}^\mathrm{des} \|_\mathrm{MSE} \; ,
    \label{eqn:loss}
\end{equation}
where~$\boldsymbol{D}$ denotes a statistical descriptor or a weighted concatenation of multiple descriptors that is computed on the reconstruction result, while~$\boldsymbol{D}^\mathrm{des}$ denotes the desired value computed from the reference structure.
Because~$m^\mathrm{end}$ results from the temporal evolution of~$f_{\boldsymbol{\theta}}$, it depends on the parameter vector~$\boldsymbol{\theta}$ of the NCA.
The central idea is train the NCA by gradient-based optimization of~$\boldsymbol{\theta}$ to minimize \autoref{eqn:loss}, whereby arbitrary descriptors can be incorporated. 
While the Gram matrices used in~\cite{mordvintsev_texture_2021} can be interpreted as a statistical descriptor, the spatial two- and three-point correlations are more common in microstructure reconstruction.
The idea of using high-dimensional, differentiable descriptors for direct microstructure reconstruction is given in~\cite{seibert_reconstructing_2021}, where an efficient and differentiable formulation of the three-point correlations is given.
As another example, a differentiable approximation to lineal path function is presented in~\cite{seibert_microstructure_2022} and a descriptor based on a hierarchical wavelet transform is given in~\cite{reck_scattering_2023}.
All these descriptors are implemented in \emph{MCRpy}~\cite{seibert_microstructure_2022}.

\subsection{Implementation}\label{sec:ncaimplementation}
The implementation of a descriptor-based NCA for microstructure reconstruction is carried out based on the code for NCA texture synthesis~\cite{mordvintsev_texture_2021} and the differentiable descriptors available in \emph{MCRpy}~\cite{seibert_microstructure_2022}.
The former code is adapted to only a single non-hidden dimension~$m(\boldsymbol{x})$ as opposed to three RGB channels.
Then, \emph{MCRpy} is used to define a loss, where different descriptors such as Gram Matrices~$G$~\cite{li_transfer_2018}, correlations~$S$~\cite{jiao_modeling_2007}, variation~$\mathcal{V}$~\cite{seibert_descriptor-based_2022} and volume fraction~$\varphi$ can be combined and weighed in a single loss function in a flexible manner. 
More information on these descriptors is given in~\cite{seibert_microstructure_2022}.
\emph{MCRpy} makes use of the automatic differentiation in \emph{TensorFlow} to compute the gradient~$\partial \mathcal{L} / \partial m$.
Then, $m$ is backpropagated through time to compute~$\partial m / \partial \boldsymbol{\theta}$ and consequently~$\partial \mathcal{L} / \partial \boldsymbol{\theta}$.

Finally, a hyperparameter study is carried out on a number of structures. 
A 12-dimensional microstructure representation (i.e. 11 hidden channels) is chosen.
Hence, the NCA has 12 output and 48 input dimensions.
With a single hidden layer of 120 neurons, the network amounts to a total of $7332$ parameters.
Further hyperparameters like the number of time steps are summarized in \autoref{tab:hyperparameters}.
\begin{table}[htpb]
\caption{Hyperparameters chosen in the present work.}\label{tab:hyperparameters}
\begin{tabular}{@{}ll@{}}
\toprule
Parameter & Value  \\
\midrule
Hidden layer size   & 120  \\
Non-hidden channels   & 1  \\
Hidden channels   & 11  \\
Activation function   & ReLU  \\
Fire rate   & $0.5$  \\
Batch size & 4 \\
Checkpointing\footnotemark[1] pool size & 1024 \\
Learning rate & $2 \cdot 10^{-3}$ \\
Rollout length probability & $\mathcal{U}(32, 64)$ \\
Gradient normalization & True \\
Overflow loss coeff & $10^{4}$ \\
Descriptors & $S, {G}, \mathcal{V}$ \\
Descriptor weights & $1, 1, 100$ \\
\bottomrule
\end{tabular}
\footnotetext[1]{This technique is outlined in~\cite{mordvintsev_texture_2021} and aims to prevent the structures from decaying for long simulation times.}
\end{table}

In order to visually compare the results of descriptor-based NCA with other methods from the literature, three open-source codes are selected from GitHub.
To represent Markov-based methods, a patch-based texture synthesis\footnote{https://github.com/anopara/texture-synthesis-nonparametric-sampling} algorithm based on~\cite{efros_image_2001,liang_real-time_2001} and a pixel-based, multi-resolution texture synthesis\footnote{https://github.com/anopara/multi-resolution-texture-synthesis} algorithm based on~\cite{de_bonet_multiresolution_1997,wei_fast_2000} are chosen.
Furthermore, \emph{MCRpy}~\cite{seibert_microstructure_2022} implements differentiable microstructure characterization and reconstruction (DMCR)~\cite{seibert_reconstructing_2021,seibert_descriptor-based_2022}.
While \emph{MCRpy} is provided by previous works of the authors, the former two methods are coded and provided by Anastasia Opara.
The authors greatly acknowledge this effort and appreciate the will to share software.

\section{Numerical experiments}\label{sec:numexp}
The microstructure evolution and the range of applicability is investigated in \autoref{sec:numexpfancy}.
These results are then compared to the literature in \autoref{sec:numexpliterature}.
Finally, the scalability of descriptor-based NCA is demonstrated in \autoref{sec:numexplarge}.
All numerical experiments are carried out on a laptop with a $12^\mathrm{th}$ Gen Intel(R) Core(TM) i7-12800H CPU at $2.40$ GHz and an Nvidia A2000 GPU with $4$ GB VRAM.

\subsection{Microstructure evolution and diversity}\label{sec:numexpfancy}
\begin{figure*}[htpb]%
\centering
\subfigure[Alloy]{\includegraphics[width=0.3\textwidth]{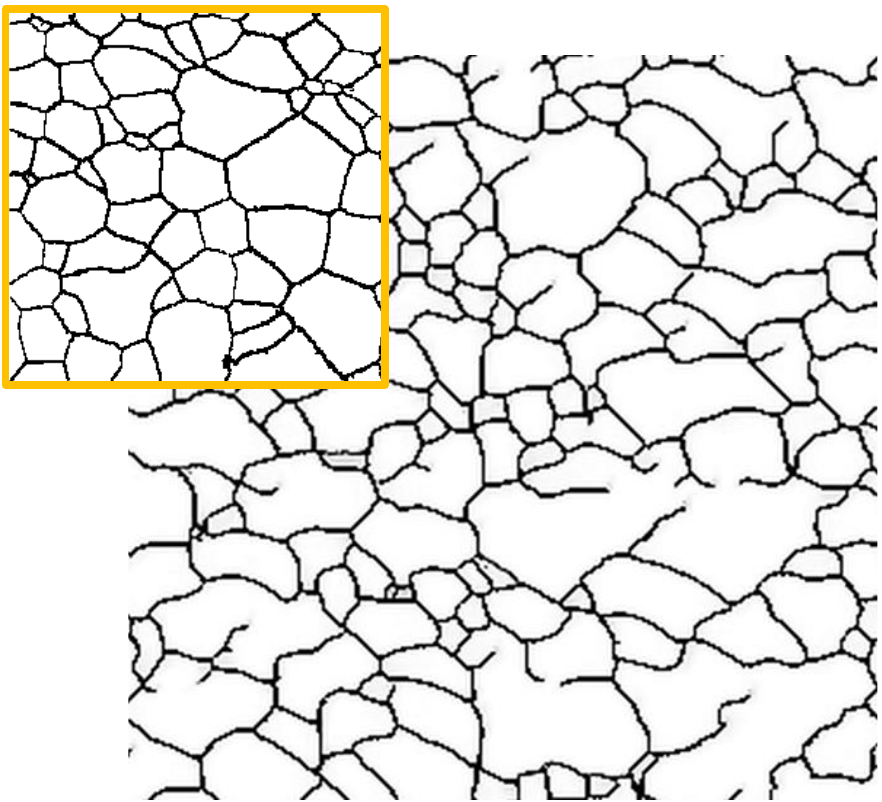}} \hfill
\subfigure[Carbonate]{\includegraphics[width=0.3\textwidth]{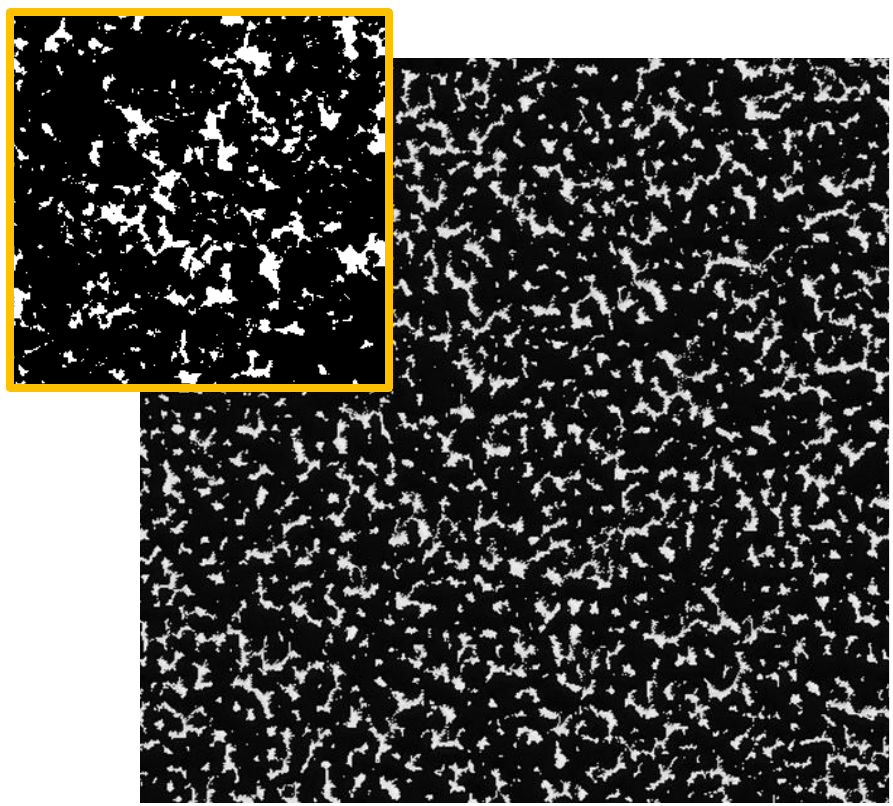}} \hfill
\subfigure[Ceramics]{\includegraphics[width=0.3\textwidth]{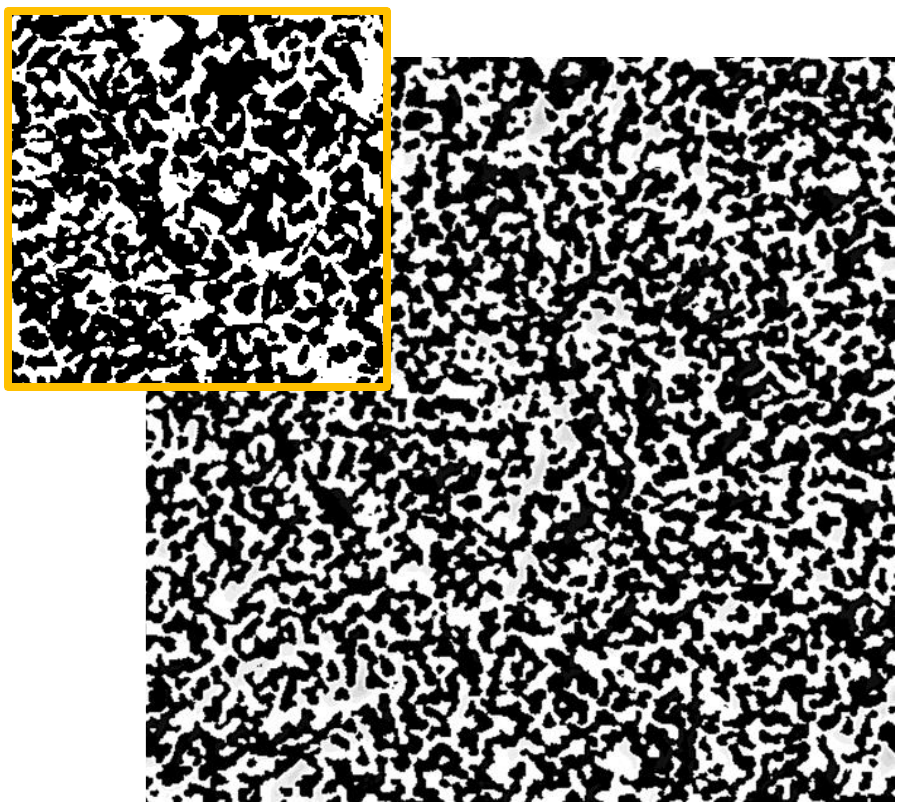}\label{fig:various_reconstructions:ceramics}} \\ 
\subfigure[Copolymer]{\includegraphics[width=0.3\textwidth]{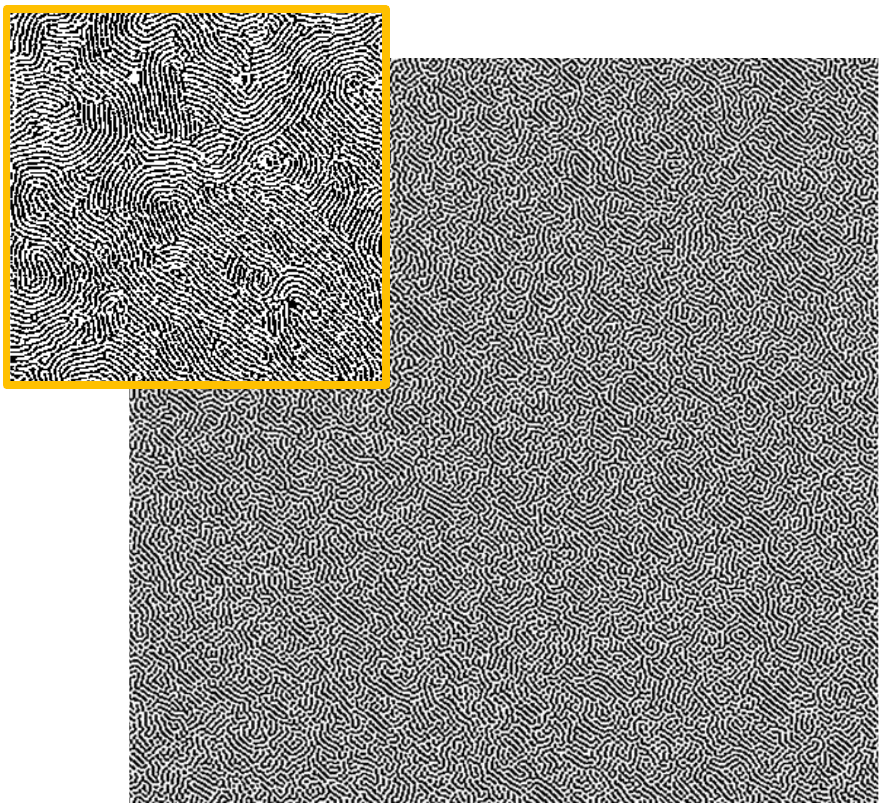}} \hfill
\subfigure[PMMA]{\includegraphics[width=0.3\textwidth]{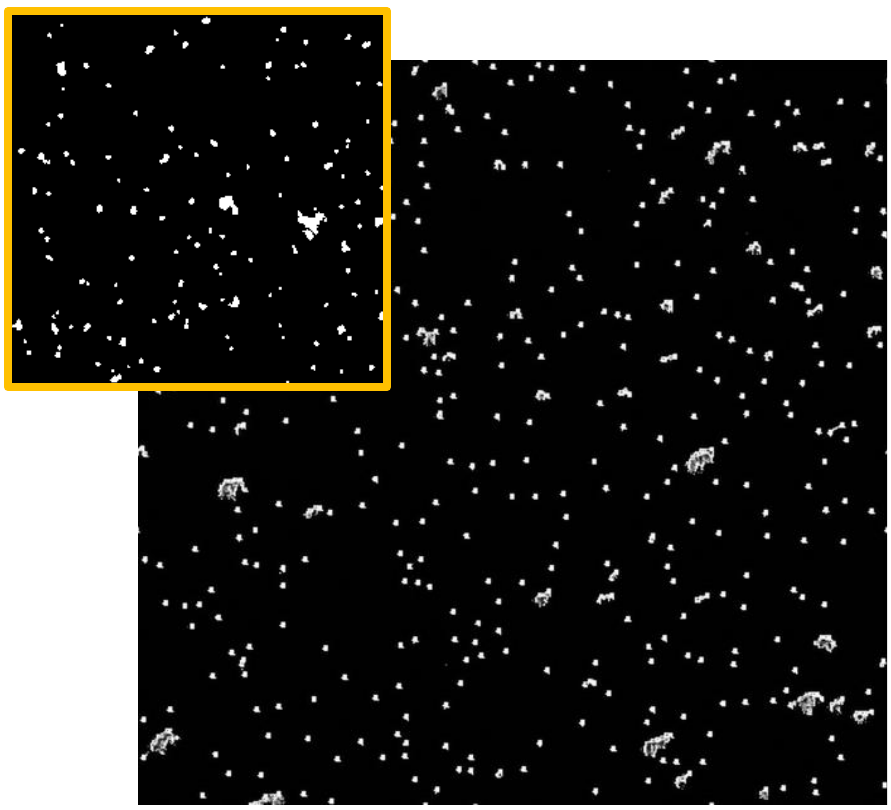}} \hfill
\subfigure[Sandstone]{\includegraphics[width=0.3\textwidth]{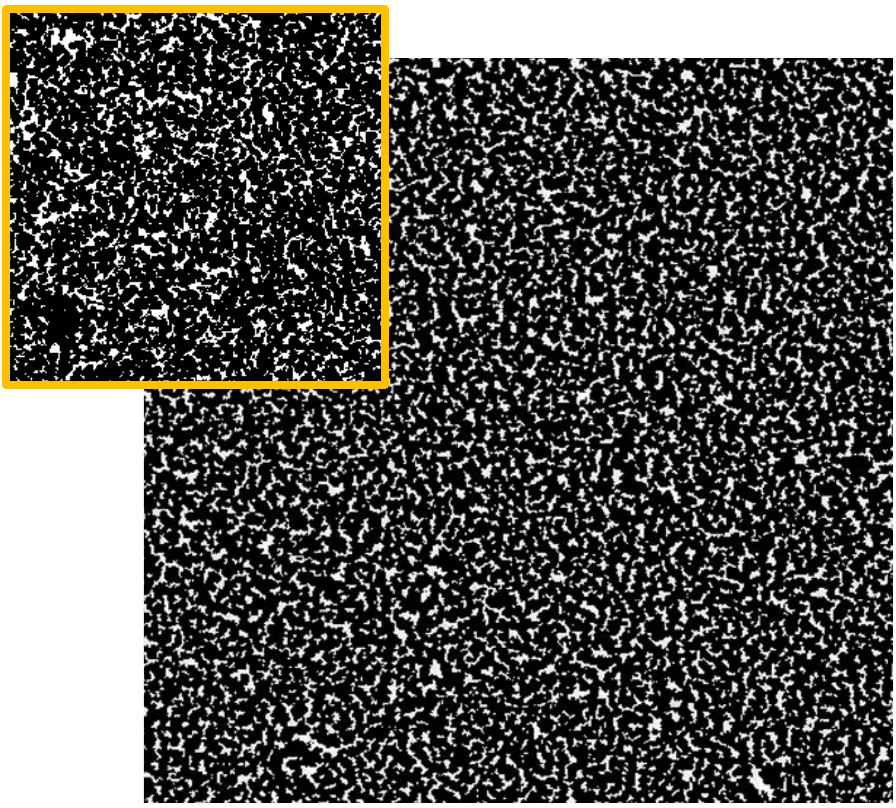}}
\caption{Reconstructions from various real materials. The original samples are given in the top left corner and are taken from~\cite{li_transfer_2018}, where they are released under the Creative Commons license~\cite{creative_commons_creative_2021}.}\label{fig:various_reconstructions}
\end{figure*}

\autoref{fig:various_reconstructions} shows reconstructions from different real materials taken from~\cite{li_transfer_2018}.
It can be seen that descriptor-based NCA are applicable to a wide variety of fundamentally different structures, ranging from relatively noise-free examples like the grain boundary structure and the ceramics to the more noisy sandstone.
Some limitations can also be seen.
As a first limitation, although the grain boundary character in the alloy is captured relatively well, not all lines are connected as in the reference case.
In order to use the results for downstream tasks like numerical simulations, a post-processing algorithm is first needed to close the gaps or eliminate unnecessary line segments.
Alternatively, it might be worth investigating if a different choice of descriptor can be used in order to better quantify the connectivity information.
Although this approach is arguably more elegant, its difficulty lies in the requirement that the descriptor should be differentiable with respect to the microstructure.
As a second limitation, it can be seen that the fingerprint-like structure of the copolymer is not adequately represented.
Although the NCA successfully creates individual sub-regions with parallel lines, these regions are not sufficiently large and the lines do not exhibit smooth curves as in the reference.
It is presently unclear to the authors how this issue can be addressed.
As a third limitation, it is noted that the probability distribtion of pixel values does not exactly match the original structures. 
Especially in the carbonate and PMMA, it can be seen that the white phase is reconstructed in bright grey color. 
Similar to the first limitation, the authors assume that a post-processing algorithm or a suitable descriptor should be sufficient to address this issue.

\begin{figure*}[htpb]%
\centering
\includegraphics[width=0.19\textwidth]{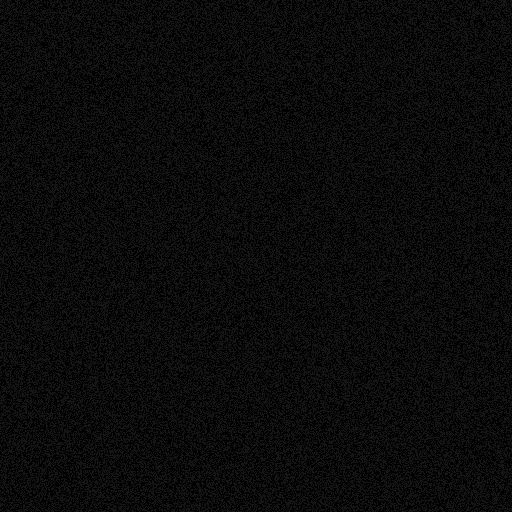} \hfill
\includegraphics[width=0.19\textwidth]{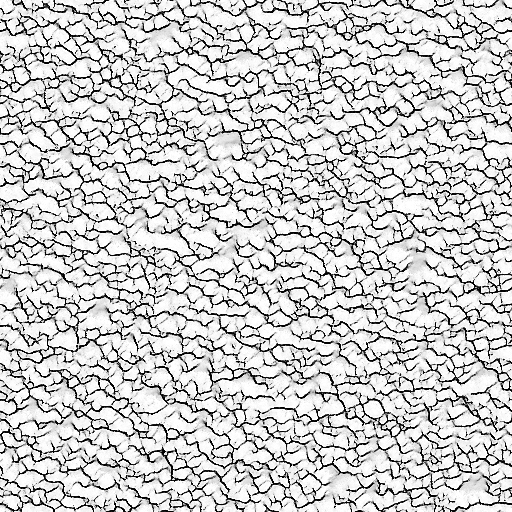} \hfill
\includegraphics[width=0.19\textwidth]{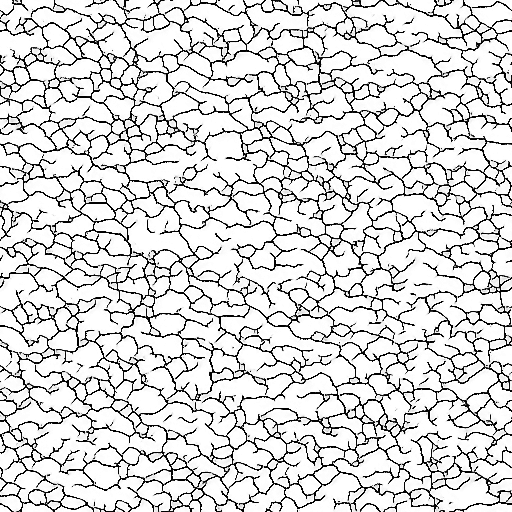} \hfill
\includegraphics[width=0.19\textwidth]{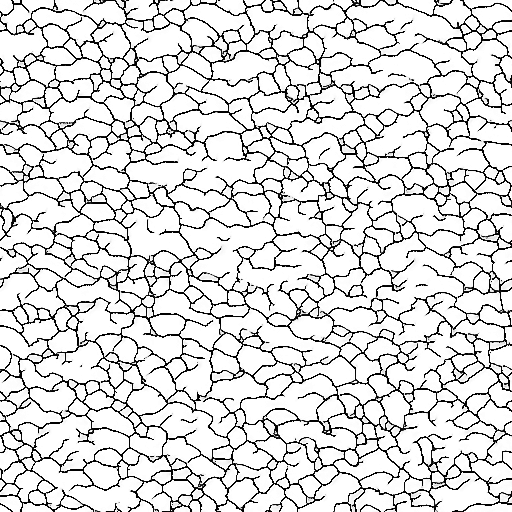} \hfill
\includegraphics[width=0.19\textwidth]{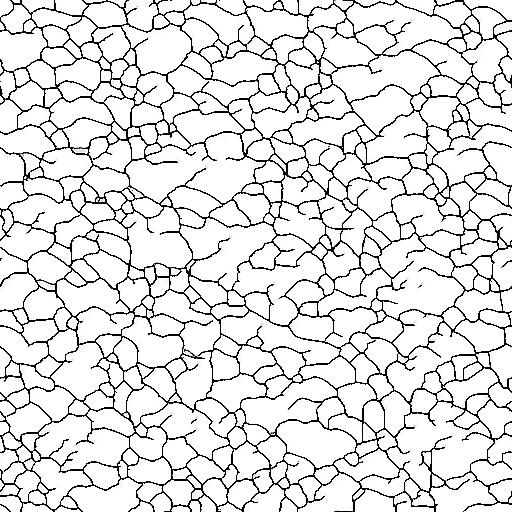} \\ \vfill
\includegraphics[width=0.19\textwidth]{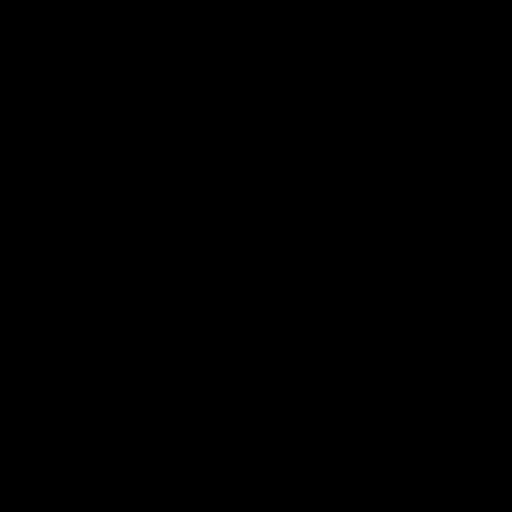} \hfill
\includegraphics[width=0.19\textwidth]{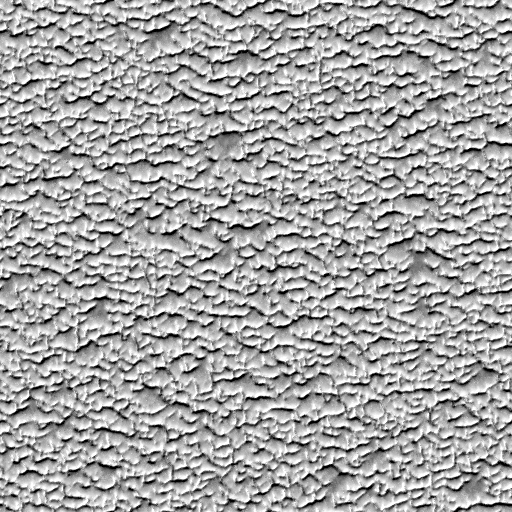} \hfill
\includegraphics[width=0.19\textwidth]{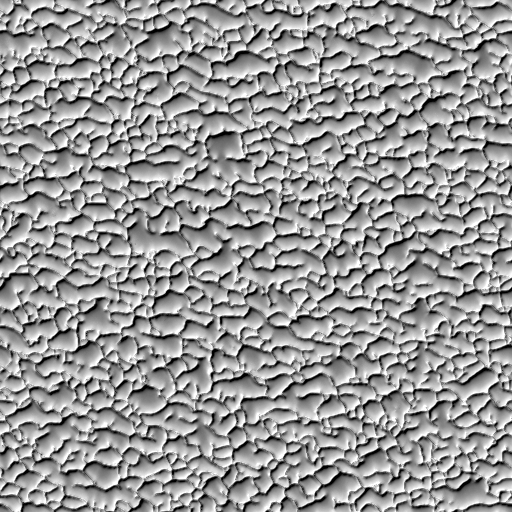} \hfill
\includegraphics[width=0.19\textwidth]{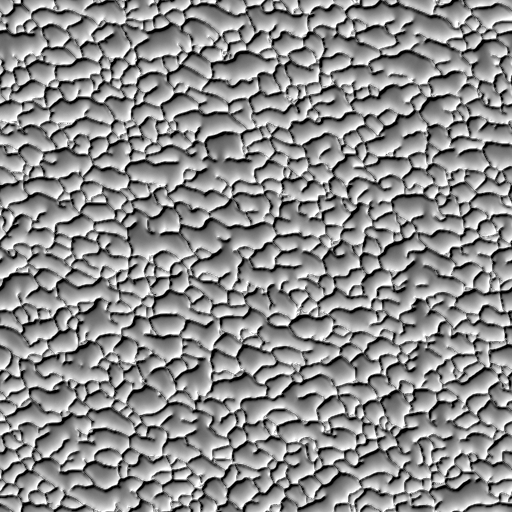} \hfill
\includegraphics[width=0.19\textwidth]{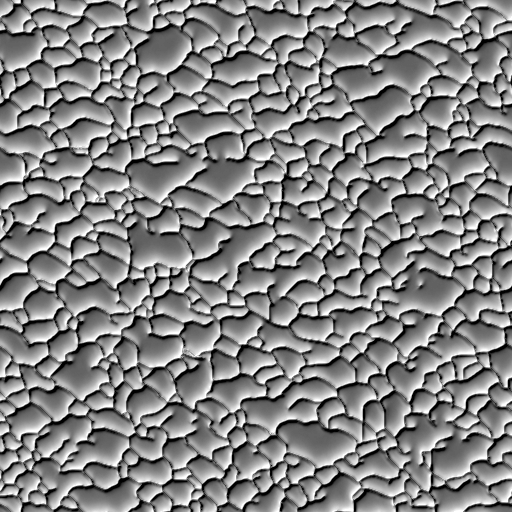} \\ \vfill
\includegraphics[width=0.19\textwidth]{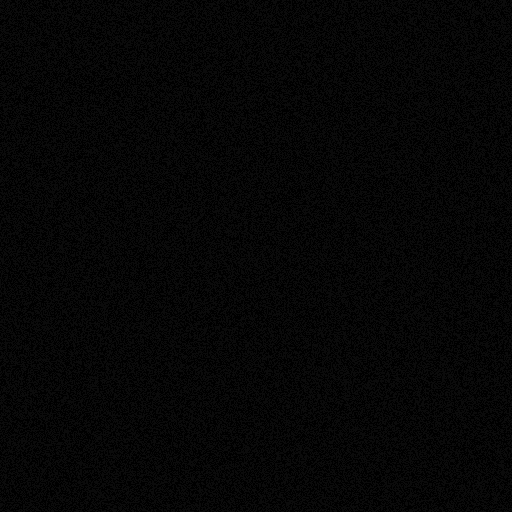} \hfill
\includegraphics[width=0.19\textwidth]{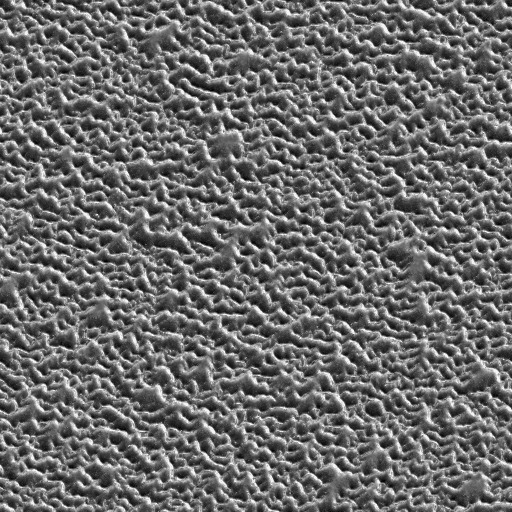} \hfill
\includegraphics[width=0.19\textwidth]{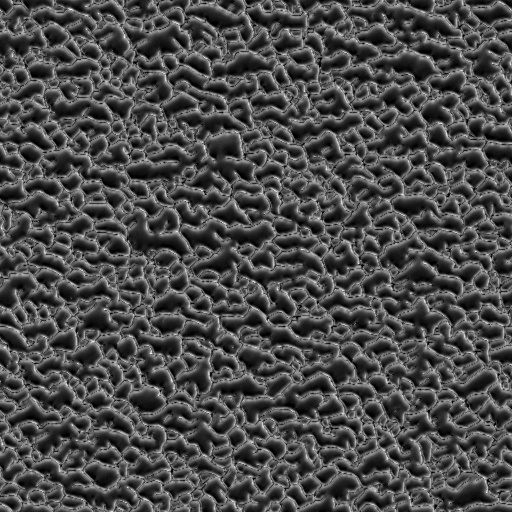} \hfill
\includegraphics[width=0.19\textwidth]{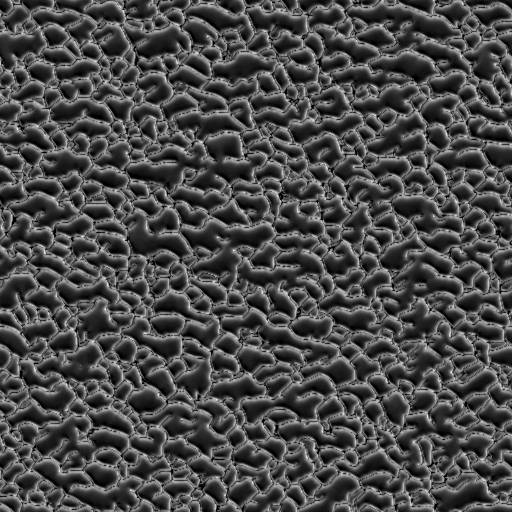} \hfill
\includegraphics[width=0.19\textwidth]{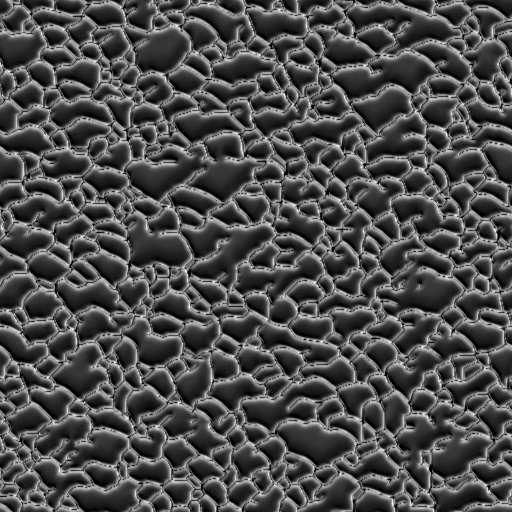} \\ \vfill
\includegraphics[width=0.19\textwidth]{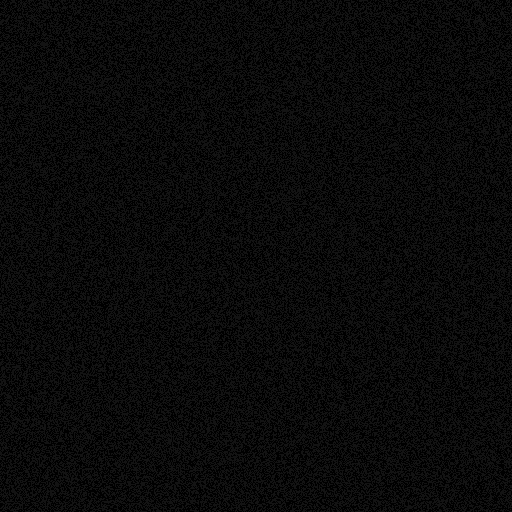} \hfill
\includegraphics[width=0.19\textwidth]{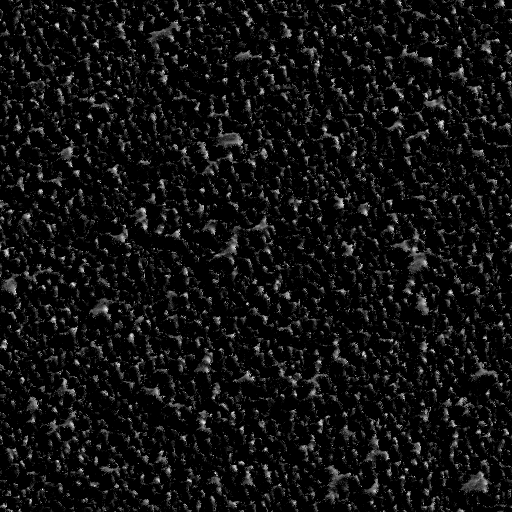} \hfill
\includegraphics[width=0.19\textwidth]{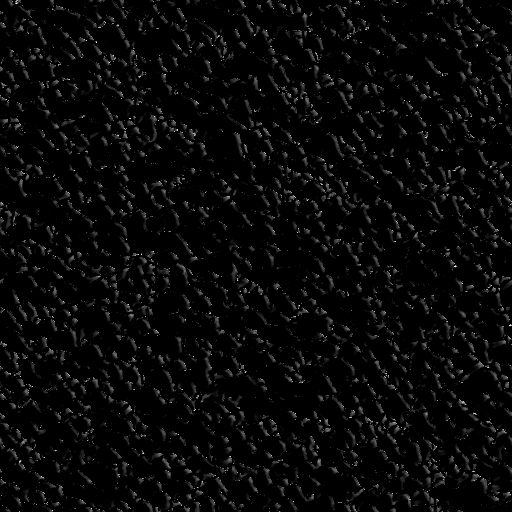} \hfill
\includegraphics[width=0.19\textwidth]{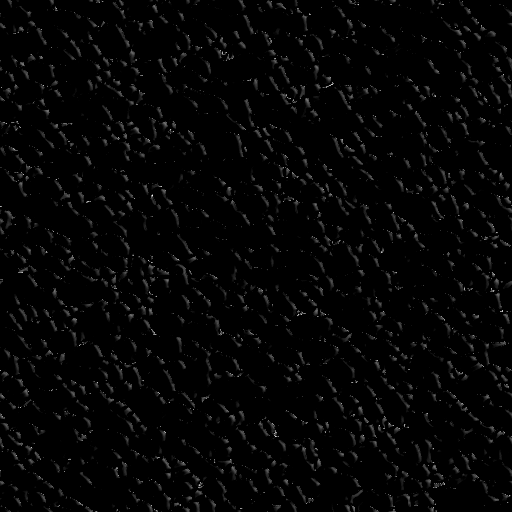} \hfill
\includegraphics[width=0.19\textwidth]{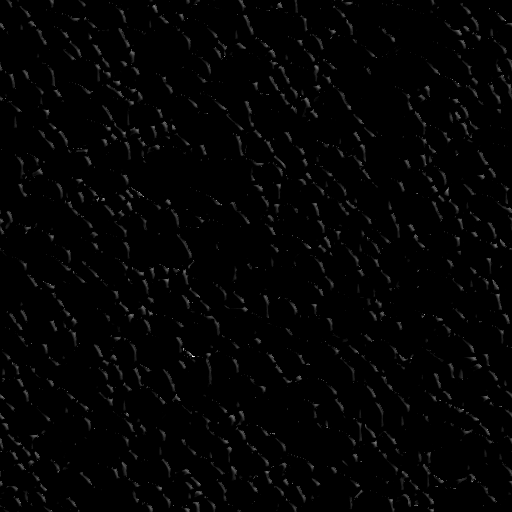} \\ \vfill
\subfigure[$t=0$]{\includegraphics[width=0.19\textwidth]{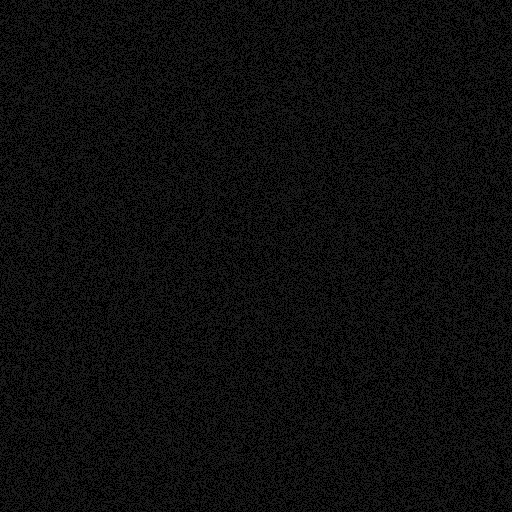}} \hfill
\subfigure[$t=30$]{\includegraphics[width=0.19\textwidth]{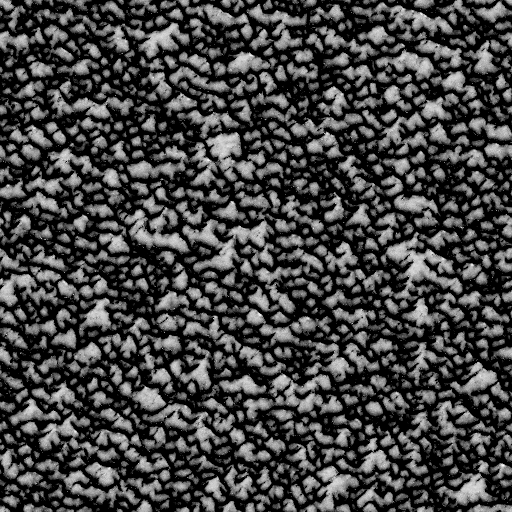}} \hfill
\subfigure[$t=60$]{\includegraphics[width=0.19\textwidth]{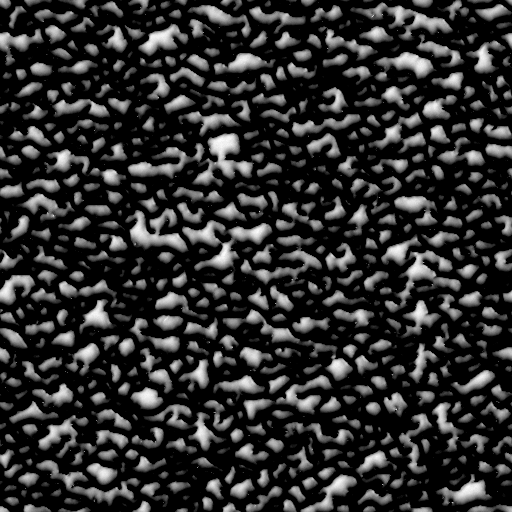}} \hfill
\subfigure[$t=90$]{\includegraphics[width=0.19\textwidth]{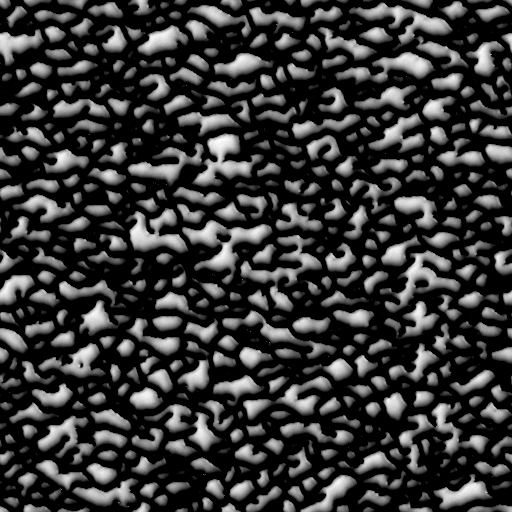}} \hfill
\subfigure[$t=300$]{\includegraphics[width=0.19\textwidth]{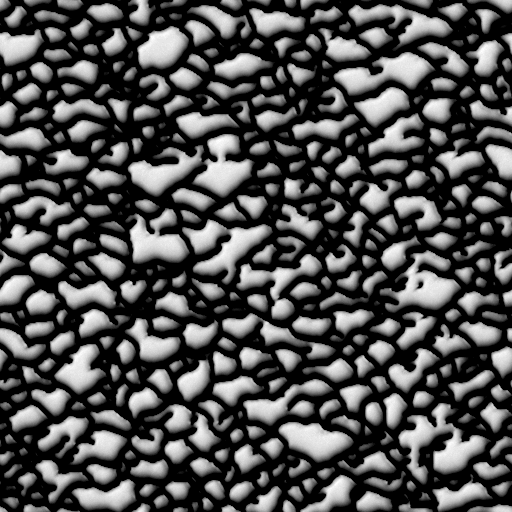}} 
\caption{The evolution of the alloy microstructure over time~$t$. The first channel is plotted in the top row and the first four hidden channels are given below. It can be seen that each hidden channel acts as a distinct feature map.}\label{fig:inter}
\end{figure*}

To provide a better understanding of the generation method, the temporal evolution of the microstructure as well as the first four hidden dimensions is plotted in \autoref{fig:inter}.
All fields are initialized by zero (black) and the structure slowly emerges.
Different hidden channels take different roles in representing structural features.
For example, the first hidden channel (second row) might be interpreted as a local vertical coordinate in each grain.
In contrast, the fourth hidden channel (last row) contains a thickened version of the gain boundaries.
Interestingly, the third hidden channel (second to last row) can  be interpreted in different ways.
It might be used as a type of residuum, since its norm decreases as the reconstruction converges.
As an alternative, it might act as a marker for specific features like triple junctions.
It can be concluded that different channels take different roles, although a direct interpretation is neither possible, nor necessary.

\begin{figure*}[htpb]%
\centering
\includegraphics[width=0.135\textwidth]{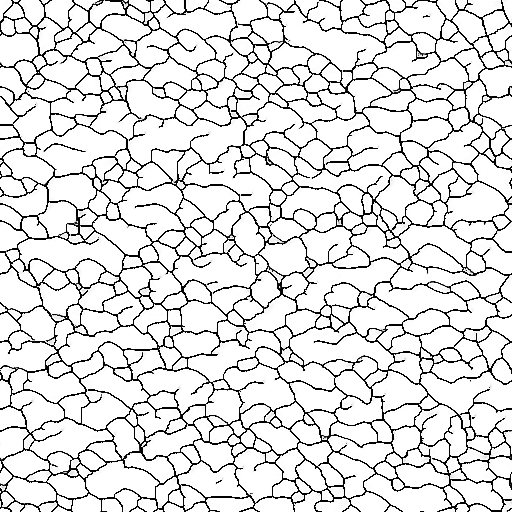} \hfill
\includegraphics[width=0.135\textwidth]{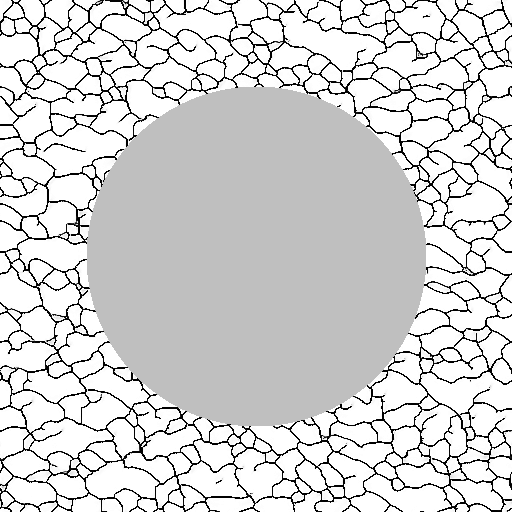} \hfill
\includegraphics[width=0.135\textwidth]{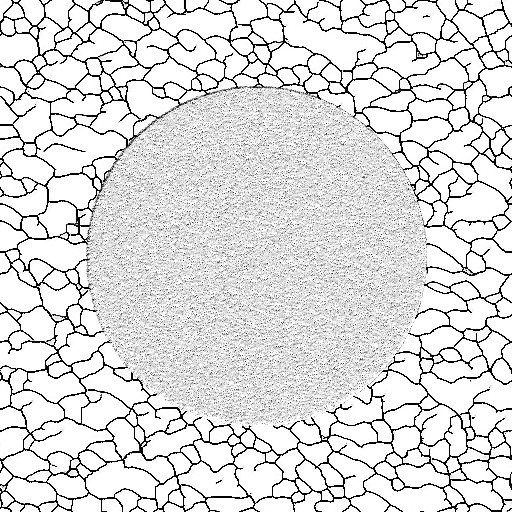} \hfill
\includegraphics[width=0.135\textwidth]{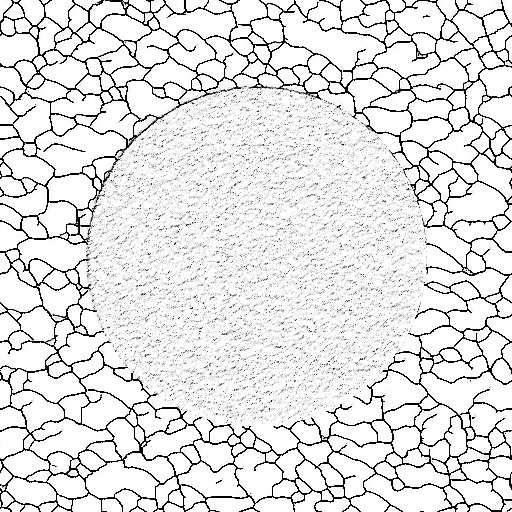} \hfill
\includegraphics[width=0.135\textwidth]{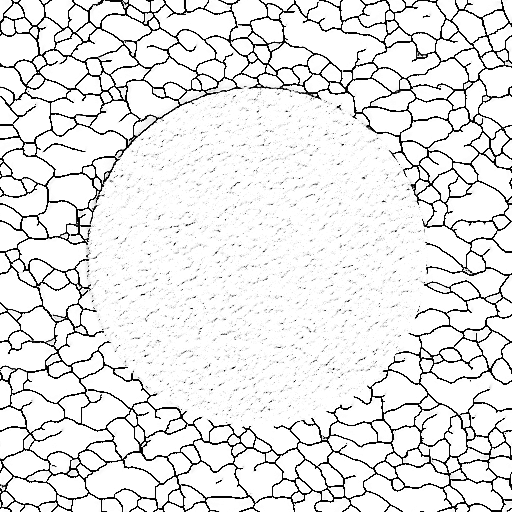} \hfill
\includegraphics[width=0.135\textwidth]{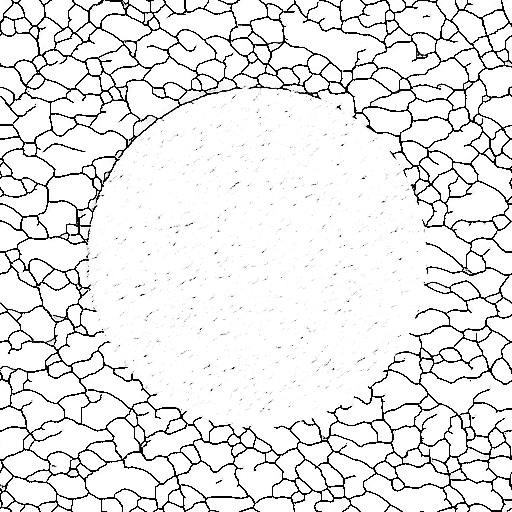} \hfill
\includegraphics[width=0.135\textwidth]{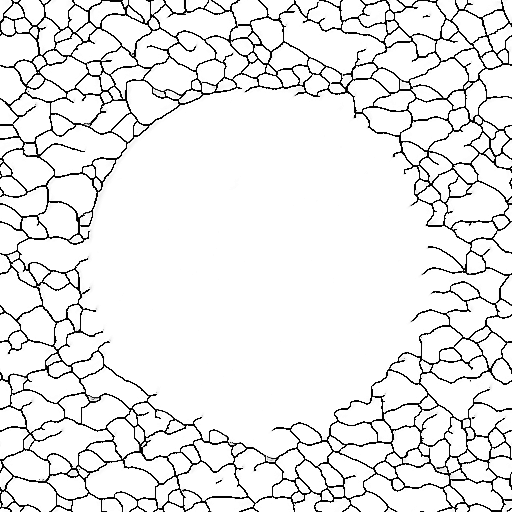} \hfill
\\ \vfill
\subfigure[$t=t'-1$]{\includegraphics[width=0.135\textwidth]{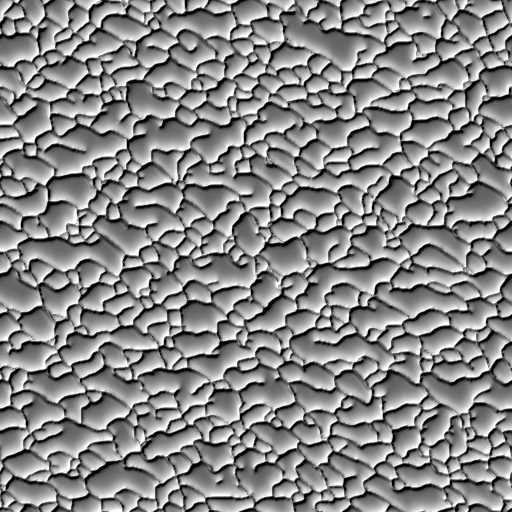}} \hfill
\subfigure[$t=t'$]{\includegraphics[width=0.135\textwidth]{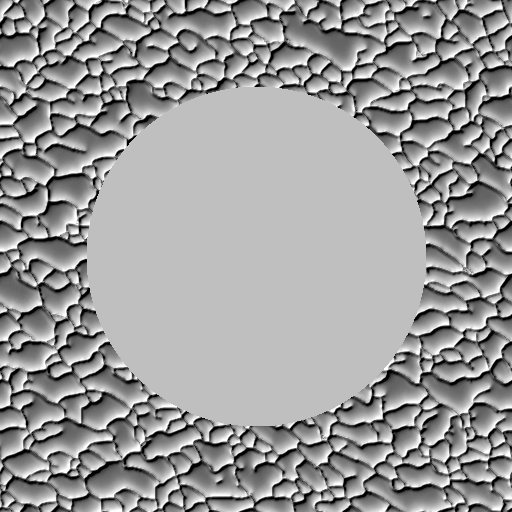}} \hfill
\subfigure[$t=t'+1$]{\includegraphics[width=0.135\textwidth]{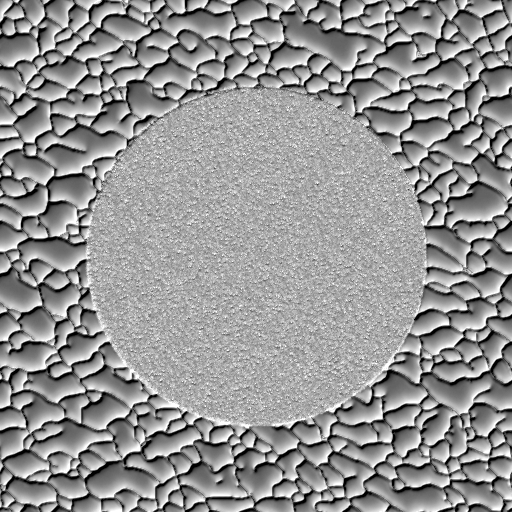}} \hfill
\subfigure[$t=t'+2$]{\includegraphics[width=0.135\textwidth]{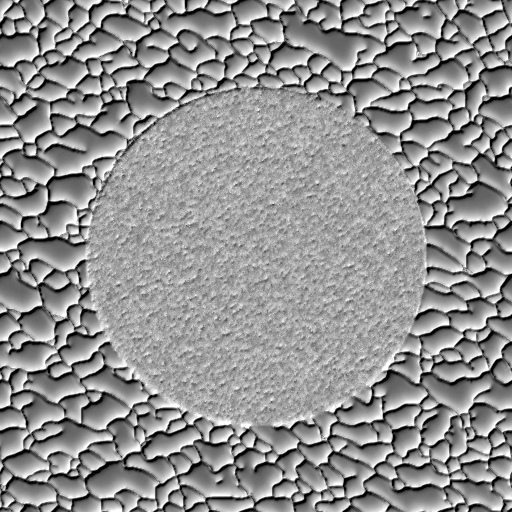}} \hfill
\subfigure[$t=t'+3$]{\includegraphics[width=0.135\textwidth]{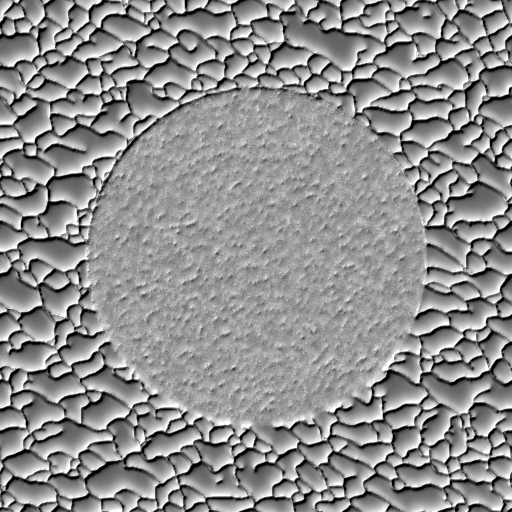}} \hfill
\subfigure[$t=t'+4$]{\includegraphics[width=0.135\textwidth]{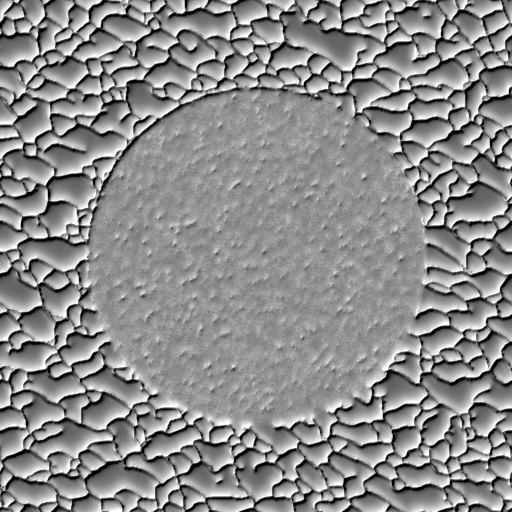}} \hfill
\subfigure[$t \gg t'$]{\includegraphics[width=0.135\textwidth]{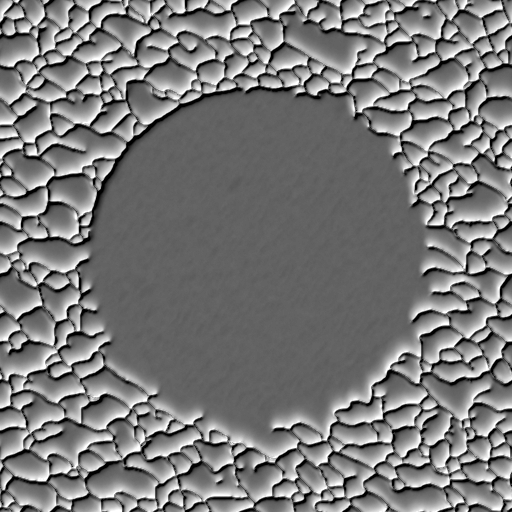}} \hfill
\\ \vfill
\caption{The role of the hidden channels is illustrated by perturbing the microstructure evolution at time~$t=t'$. All pixel values within a given radius are set to~$0.5$. In the presented case, all channels are perturbed, whereas in \autoref{fig:perturb_zero}, the hidden channels remain intact. Only the microstructure (top) and the first hidden channel (bottom) are plotted for brevity. Unlike in \autoref{fig:perturb_zero}, the structure does not recover.}\label{fig:perturb_all}
\end{figure*}

\begin{figure*}[htpb]%
\centering
\includegraphics[width=0.135\textwidth]{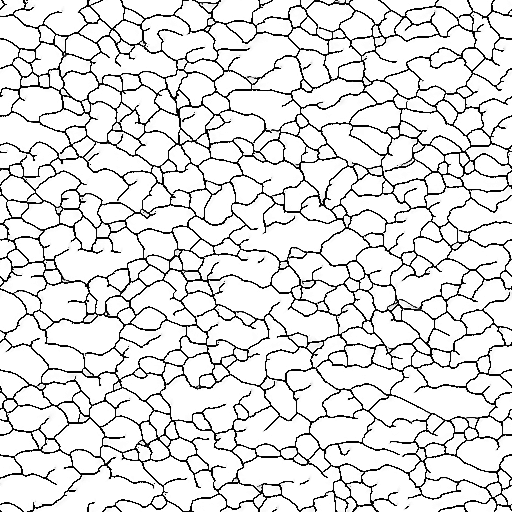} \hfill
\includegraphics[width=0.135\textwidth]{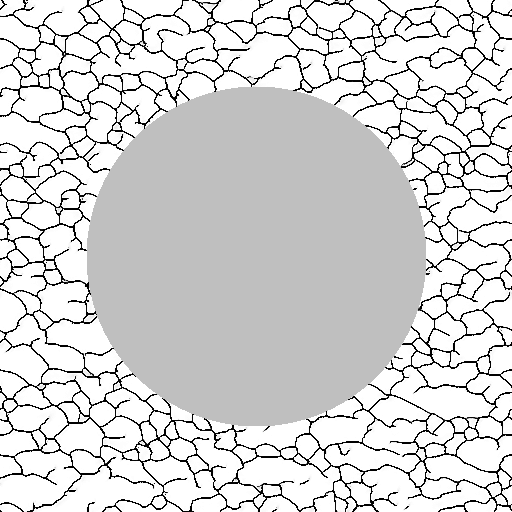} \hfill
\includegraphics[width=0.135\textwidth]{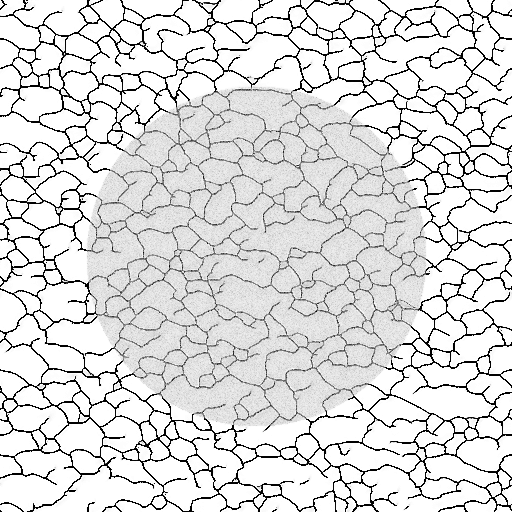} \hfill
\includegraphics[width=0.135\textwidth]{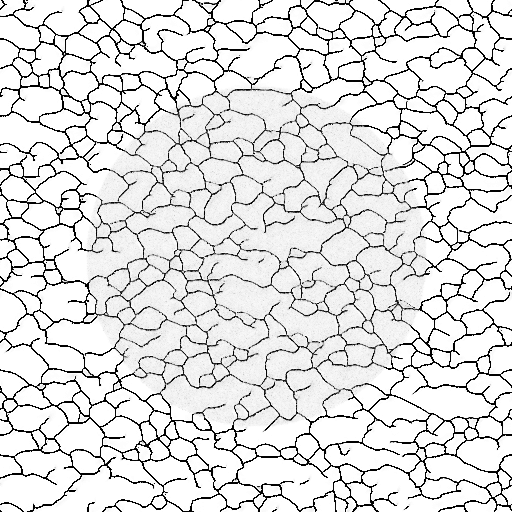} \hfill
\includegraphics[width=0.135\textwidth]{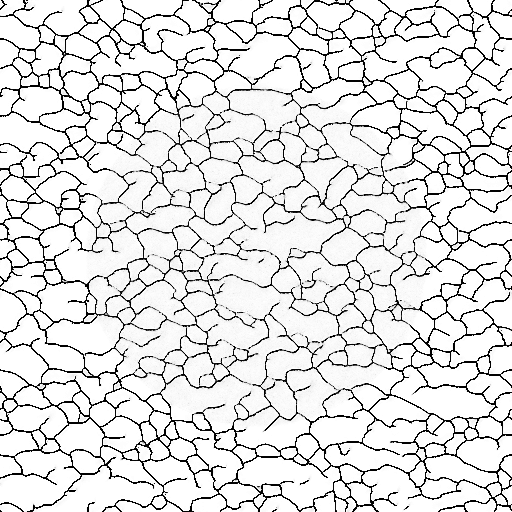} \hfill
\includegraphics[width=0.135\textwidth]{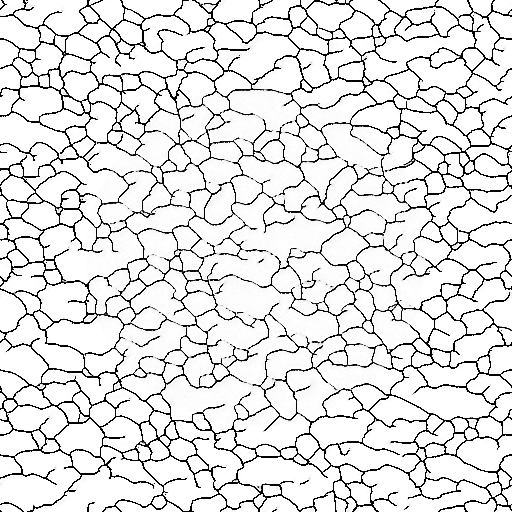} \hfill
\includegraphics[width=0.135\textwidth]{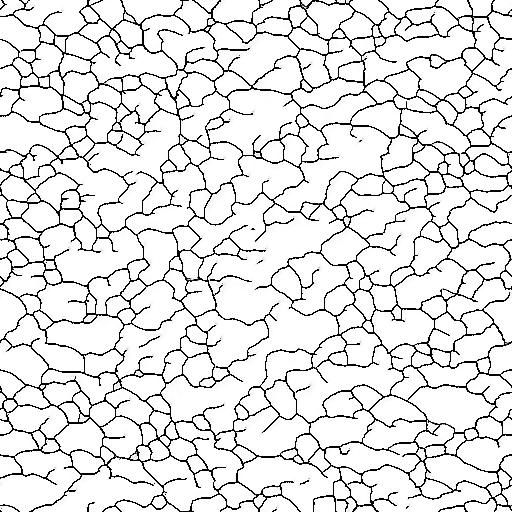} \hfill
\\ \vfill
\subfigure[$t=t'-1$]{\includegraphics[width=0.135\textwidth]{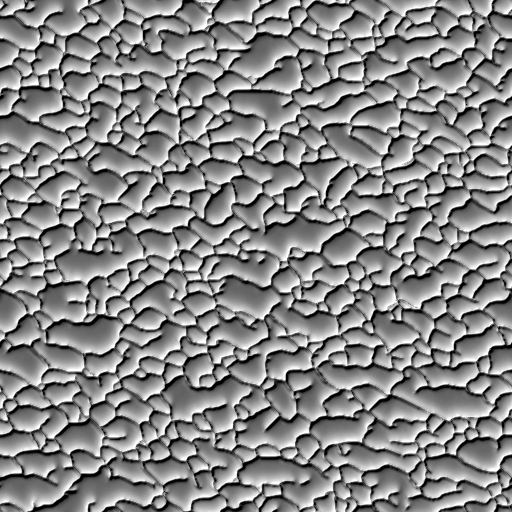}} \hfill
\subfigure[$t=t'$]{\includegraphics[width=0.135\textwidth]{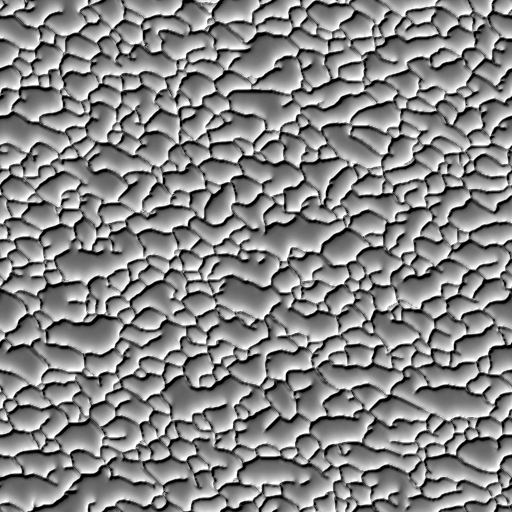}} \hfill
\subfigure[$t=t'+1$]{\includegraphics[width=0.135\textwidth]{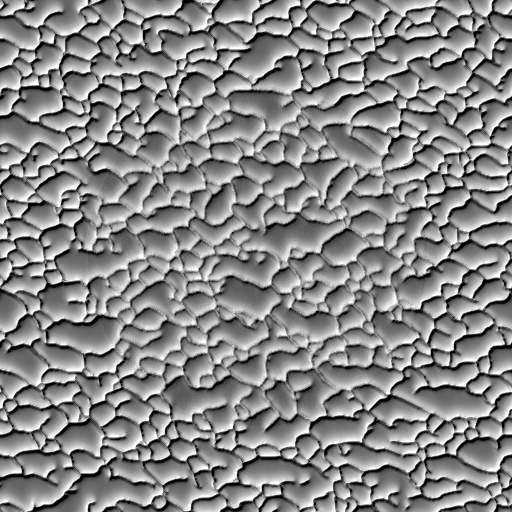}} \hfill
\subfigure[$t=t'+2$]{\includegraphics[width=0.135\textwidth]{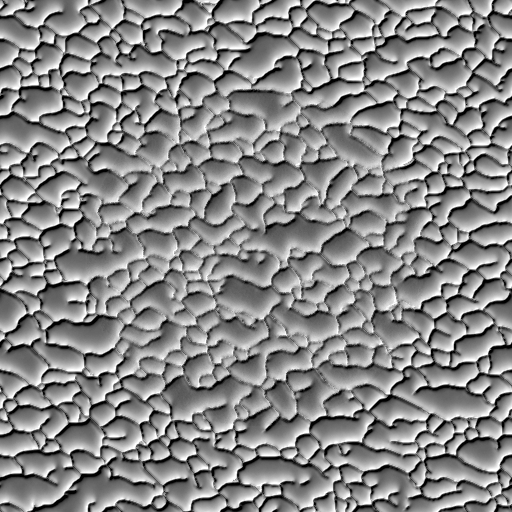}} \hfill
\subfigure[$t=t'+3$]{\includegraphics[width=0.135\textwidth]{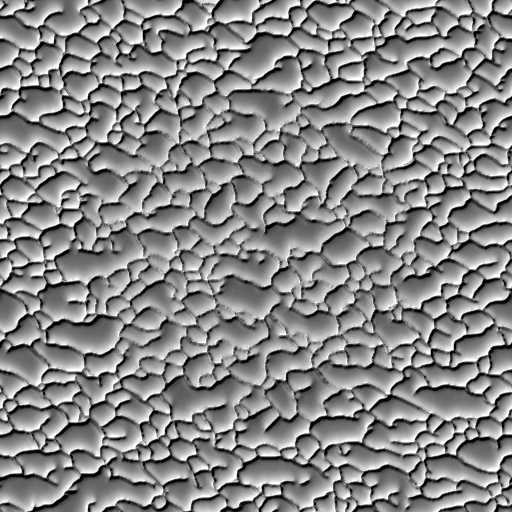}} \hfill
\subfigure[$t=t'+4$]{\includegraphics[width=0.135\textwidth]{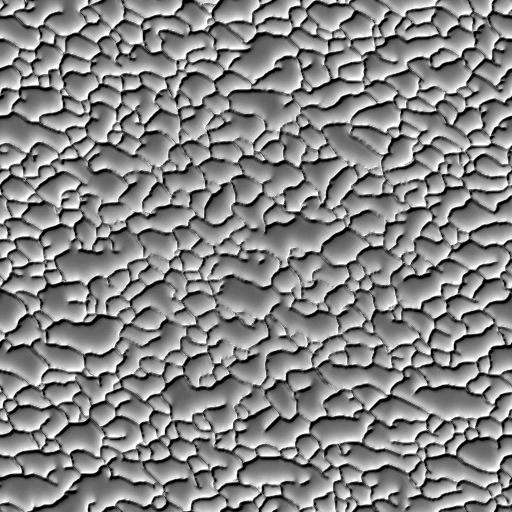}} \hfill
\subfigure[$t \gg t'$]{\includegraphics[width=0.135\textwidth]{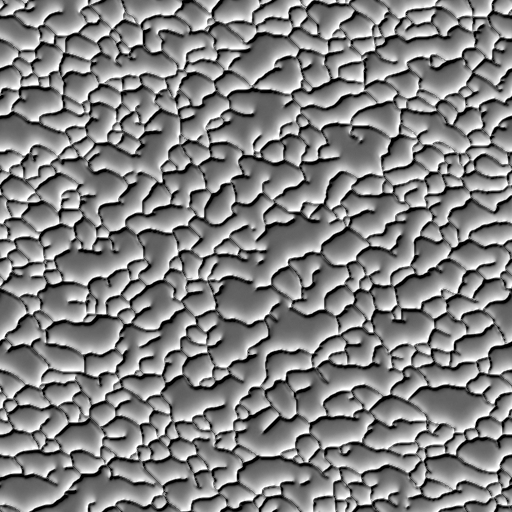}} \hfill
\\ \vfill
\caption{The role of the hidden channels is illustrated by perturbing the microstructure evolution at time~$t=t'$. All pixel values within a given radius are set to~$0.5$. In the presented case, the hidden channels remain intact, whereas in \autoref{fig:perturb_all}, all channels are perturbed. Only the microstructure (top) and the first hidden channel (bottom) are plotted for brevity. Unlike in \autoref{fig:perturb_all}, the structure recovers, albeit to a different solution.}\label{fig:perturb_zero}
\end{figure*}

It is demonstrated in the works of Mordvintsev et al.~\cite{mordvintsev_growing_2020,mordvintsev_texture_2021} that the NCA-based generation process is often robust with respect to perturbations. 
To test whether this trend is transferred to descriptor-based NCA for microstructure reconstruction, two numerical experiments are carried out.
After the generation process has converged to a good solution, the structure is perturbed by setting all values within a circular radius to $0.5$.
This is applied to all channels in \autoref{fig:perturb_all} and only to the non-hidden dimension in \autoref{fig:perturb_zero}.
It can be seen that the structure only recovers in the latter case.
Besides stressing the key role of the hidden channels, this indicates that the robustness of NCA is only partially observed in descriptor-based NCA for microstructure reconstruction.

\subsection{Comparison to literature}\label{sec:numexpliterature}
\begin{figure*}[htpb]%
\centering
\includegraphics[width=0.19\textwidth]{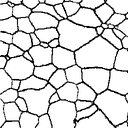} 
\includegraphics[width=0.19\textwidth]{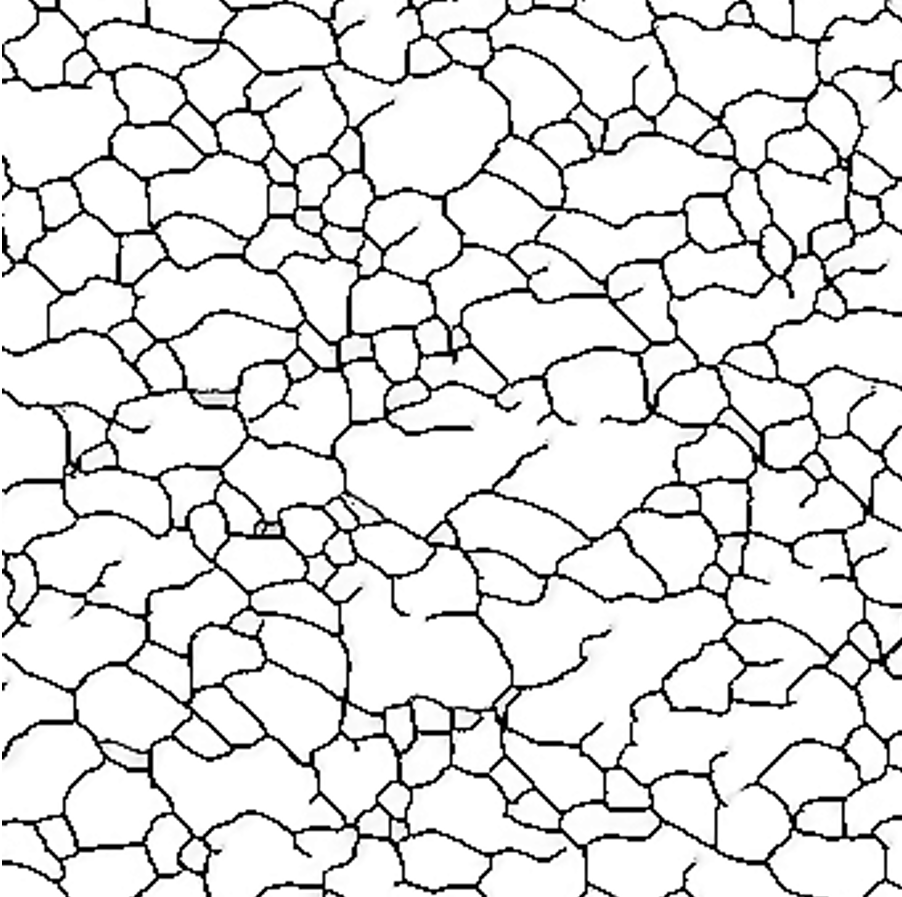}
\includegraphics[width=0.19\textwidth]{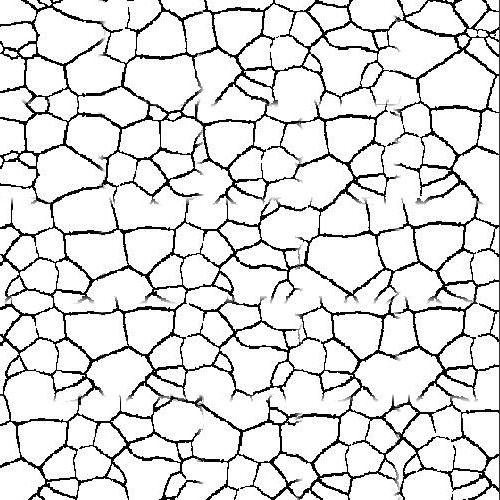}
\includegraphics[width=0.19\textwidth]{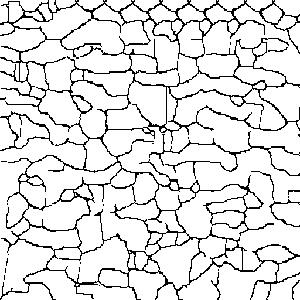} 
\includegraphics[width=0.19\textwidth]{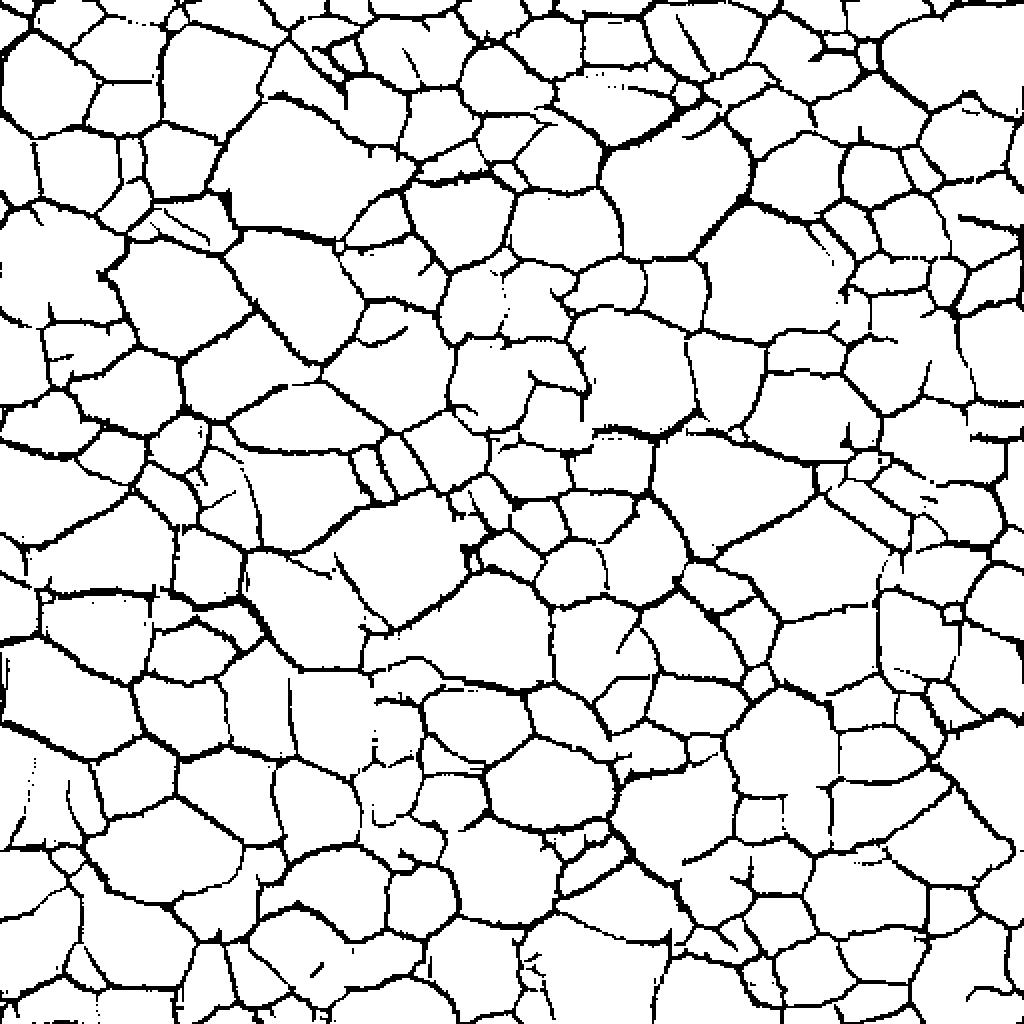} \\ \vspace{0.3cm}
\includegraphics[width=0.19\textwidth]{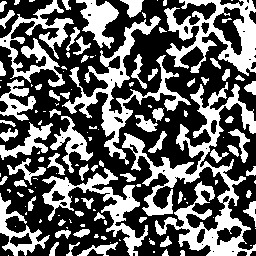} 
\includegraphics[width=0.19\textwidth]{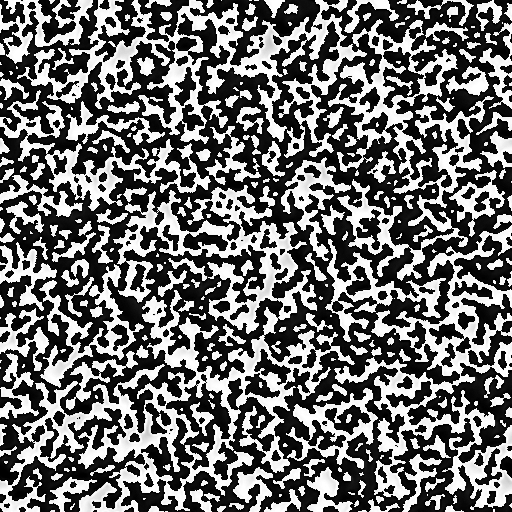}
\includegraphics[width=0.19\textwidth]{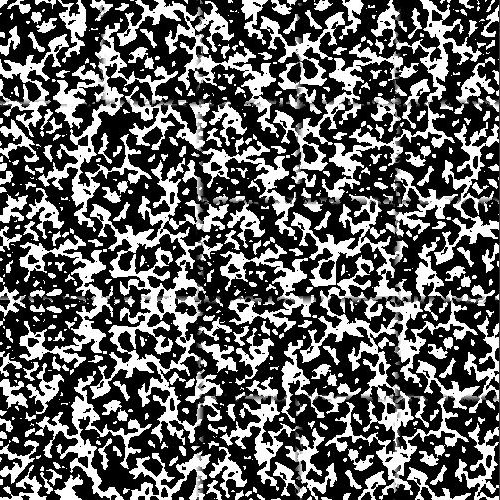}
\includegraphics[width=0.19\textwidth]{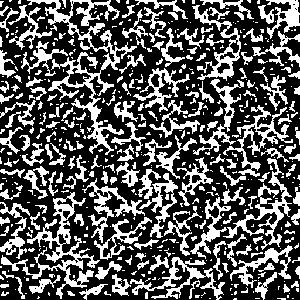} 
\includegraphics[width=0.19\textwidth]{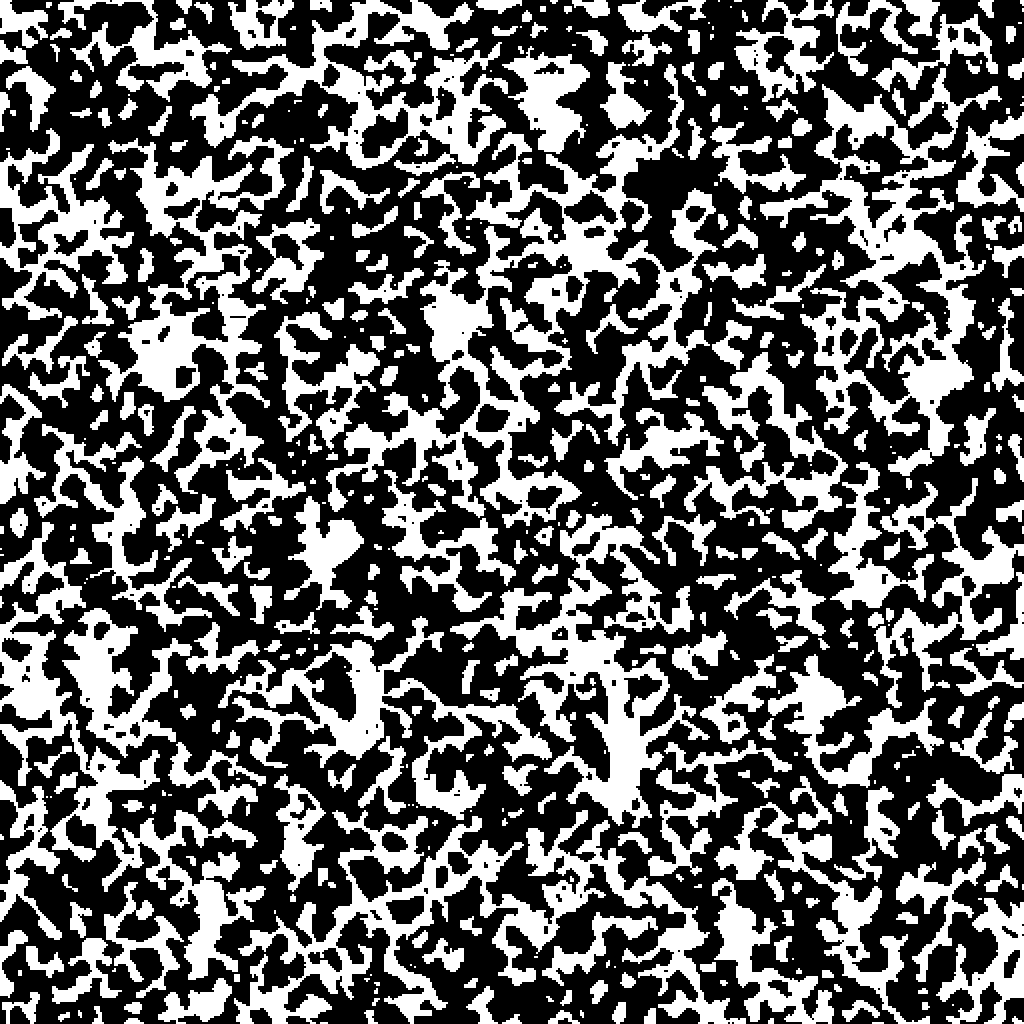} \\
\subfigure[Reference]{\includegraphics[width=0.19\textwidth]{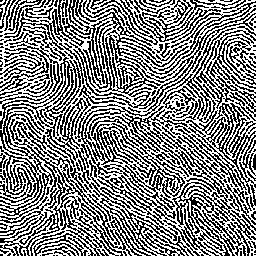}} 
\subfigure[NCA]{\includegraphics[width=0.19\textwidth]{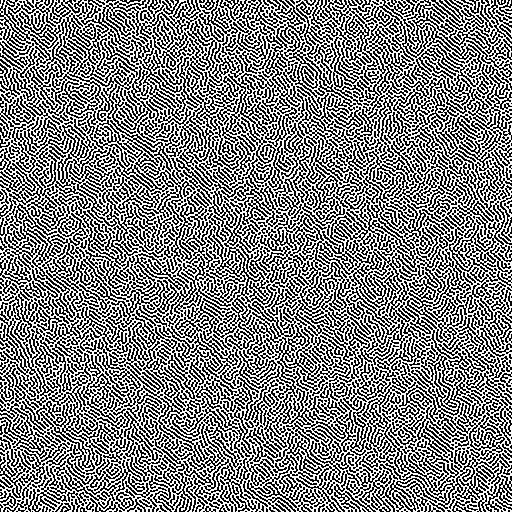}} 
\subfigure[PBTS]{\includegraphics[width=0.19\textwidth]{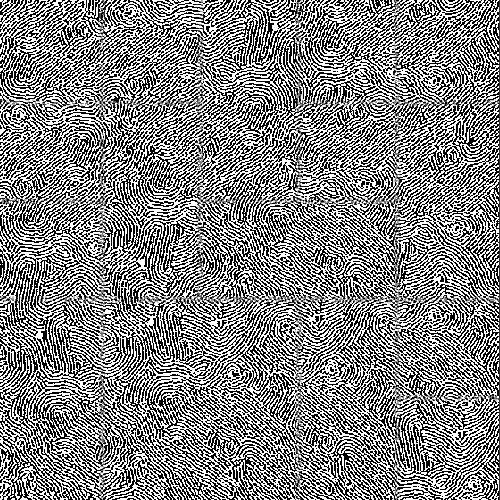}} 
\subfigure[MRTS]{\includegraphics[width=0.19\textwidth]{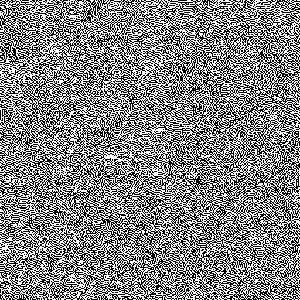}} 
\subfigure[DMCR]{\includegraphics[width=0.19\textwidth]{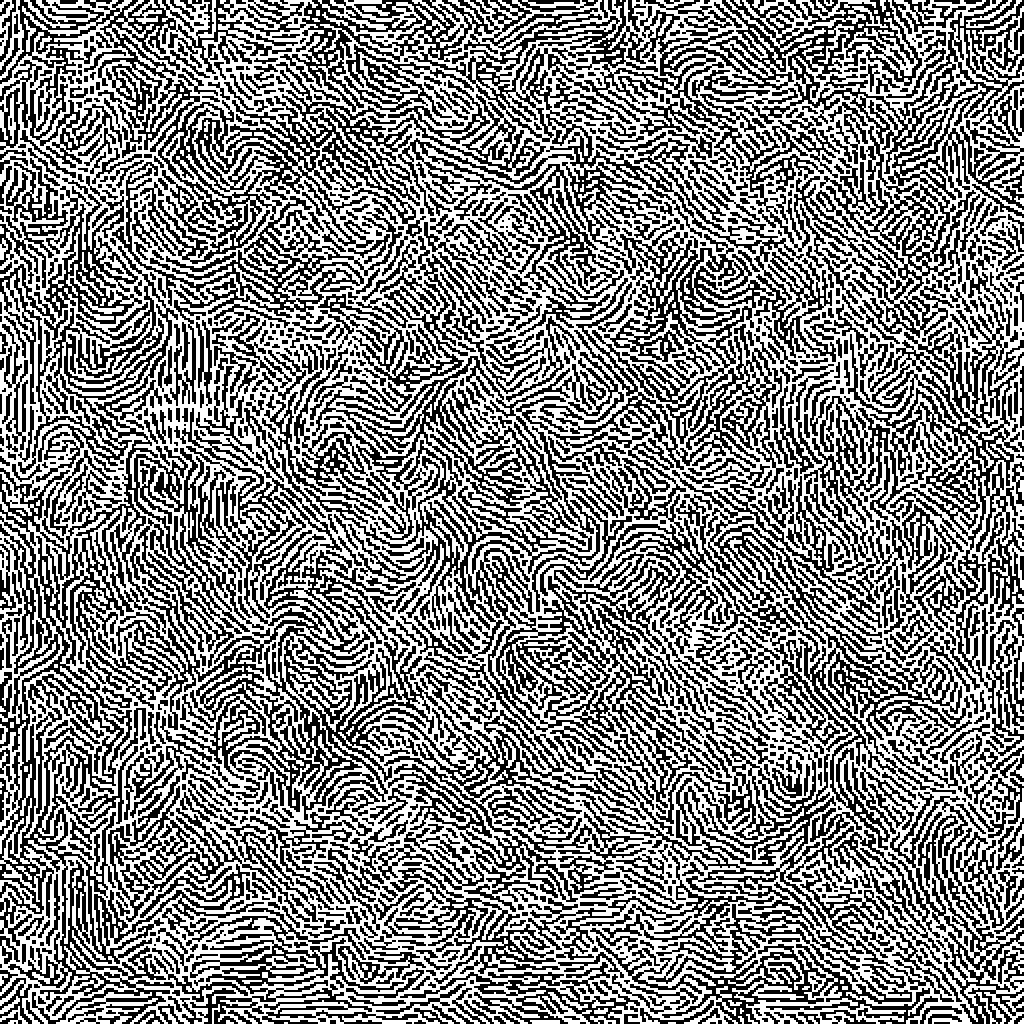}} \\
\caption{Comparison of three selected materials reconstructions with methods from the literature. Patch-based texture synthesis (PBTS) and multi-resolution texture synthesis are Markov-based approaches, while differentiable microstructure characterization and reconstruction (DMCR) is descriptor-based. The reconstructed structures are two times larger than the reference in each direction. More information is given in \autoref{sec:ncaimplementation}.}\label{fig:comparison_literature}
\end{figure*}

In order to compare descriptor-based NCA reconstruction results to the literature, two Markov- and one descriptor-based approach is chosen\footnote{
The authors were not able to apply deep learning-based algorithms, since only a single reference sample is available.
Similarly, simulation-based approaches are omitted because they require a detailed analysis and modeling of the materials' underlying physics that exceeds the scope of this work. 
}.
\autoref{fig:comparison_literature} shows the results, where only three material classes are selected for the sake of brevity. 
At a first glance, all methods produce high-quality results.
Patch-based texture synthesis, however, does not produce new structural features, but copies pathes from the original structure to different locations in the target image.
The patch boundaries can be distinguished upon closer inspection.
As a pixel-based approach, multi-resolution texture synthesis does not suffer from this phenomenon.
However, especially in the alloy, strange features like completely vertical grain boundaries can be observed and the structure coincidentally repeats itself in the top right corner.
Furthermore, the highly complex fingerprint-like copolymer structure is not captured adequately.
Finally, DMCR as a descriptor-based method\footnote{The Yeong-Torquato algorithm can be expected to yield equally good or even better results, however, at a significantly higher computational cost.} produces good results for all considered materials.
While the alloy and ceramics are similarly well reconstructed as with the descriptor-based NCA, DMCR produces visually superior results for the copolymer.
It can be concluded that the result quality of NCA outperforms standard Markov-based techniques and almost reaches that of direct descriptor-based optimization.
The advantage, with respect to the latter lies in the performance and scalability as discussed in the following.

\subsection{Performance and scalability}\label{sec:numexplarge}
An objective assessment of the reconstruction results in terms of microstructure descriptors is paramount to evaluating the accuracy of any reconstruction algorithm.
In this context, it should be mentioned that the presented method is only partially descriptor-based since the descriptors are used during training, but not during sampling.
For this reason, independent realizations of the material exhibit random deviations from the target descriptor.
Naturally, these fluctuations are expected to decrease as the microstructure size increases.

\begin{figure*}[htpb]%
\centering
\includegraphics[width=0.16\textwidth]{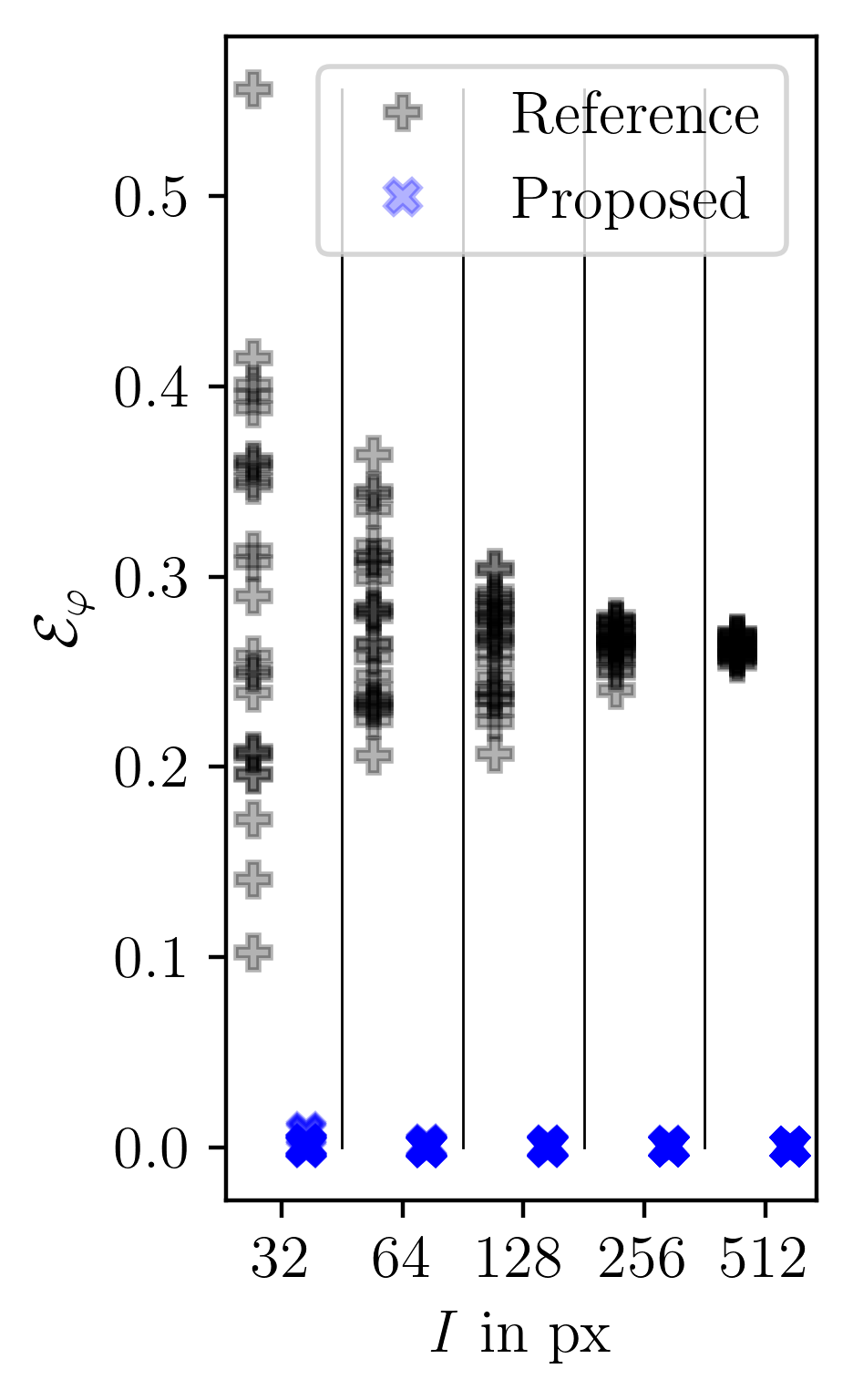}
\includegraphics[width=0.16\textwidth]{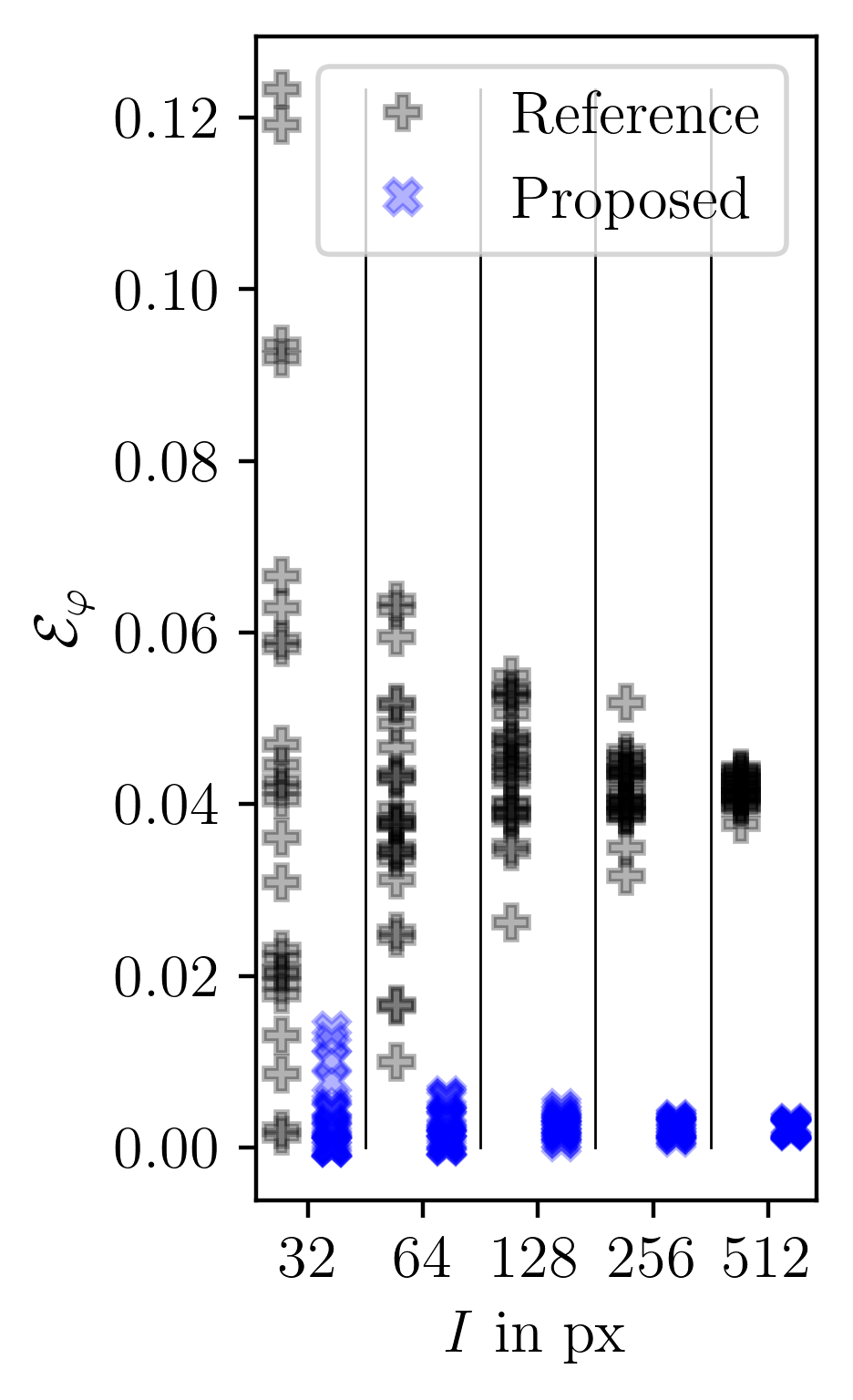}
\includegraphics[width=0.16\textwidth]{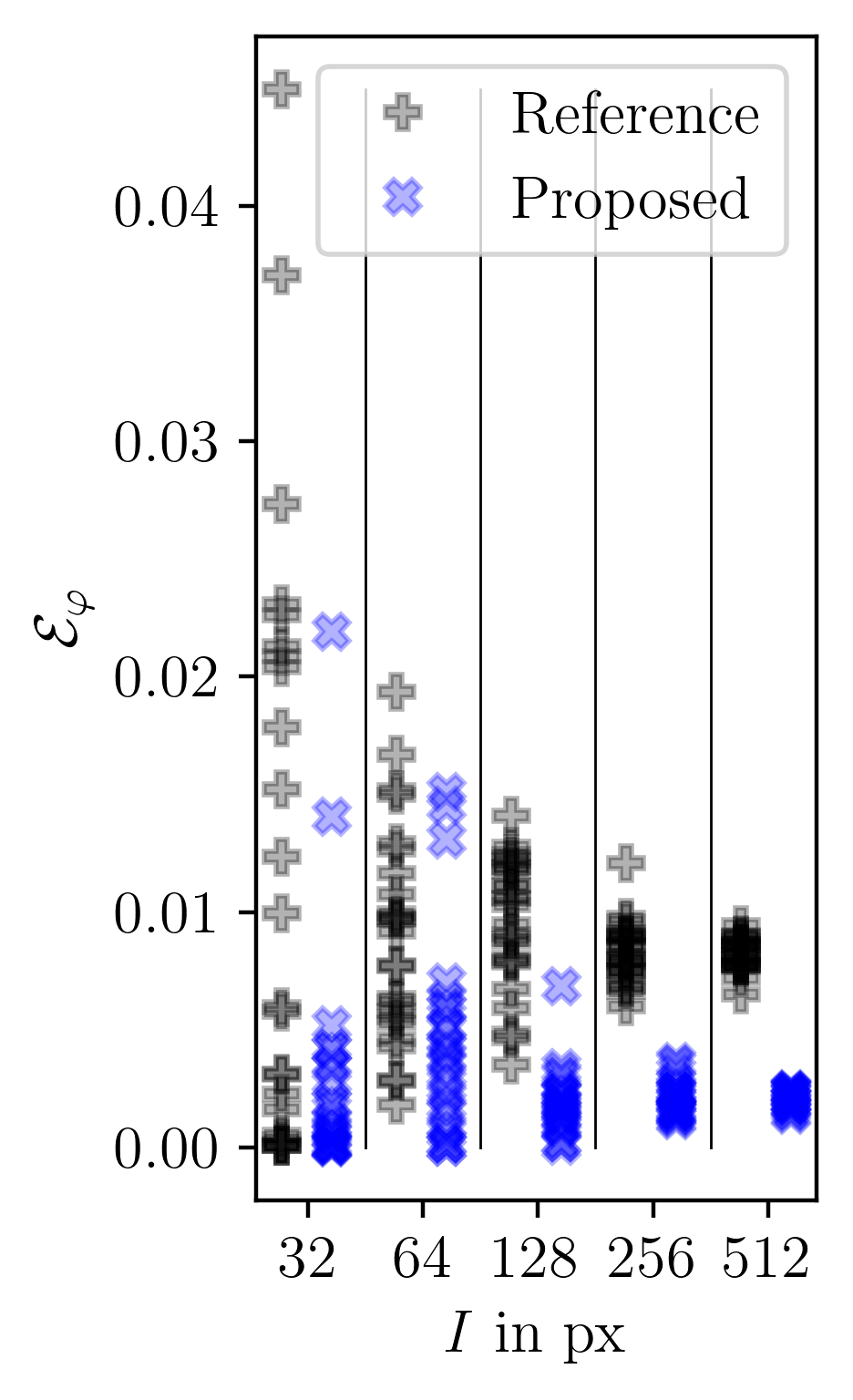}
\includegraphics[width=0.16\textwidth]{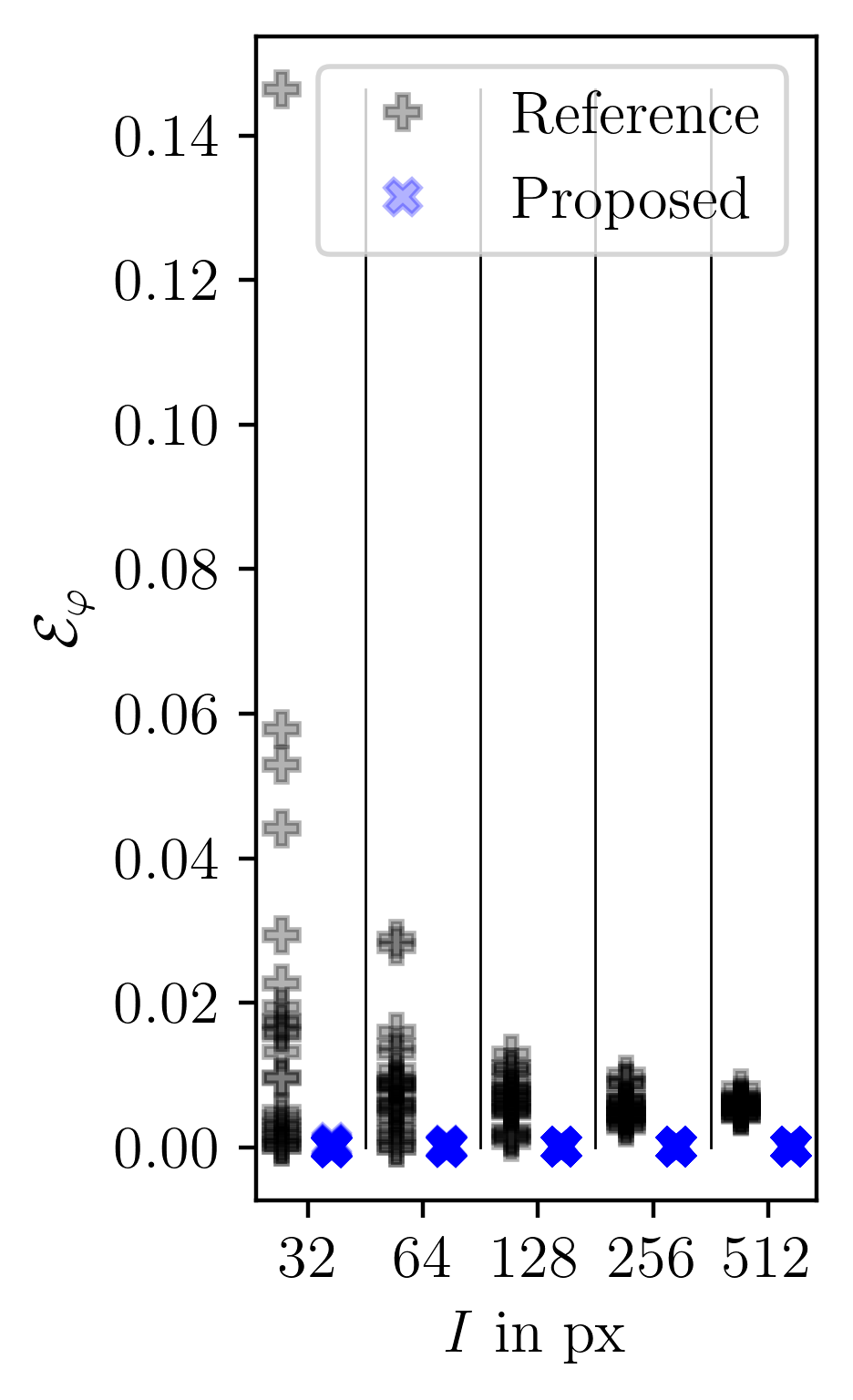}
\includegraphics[width=0.16\textwidth]{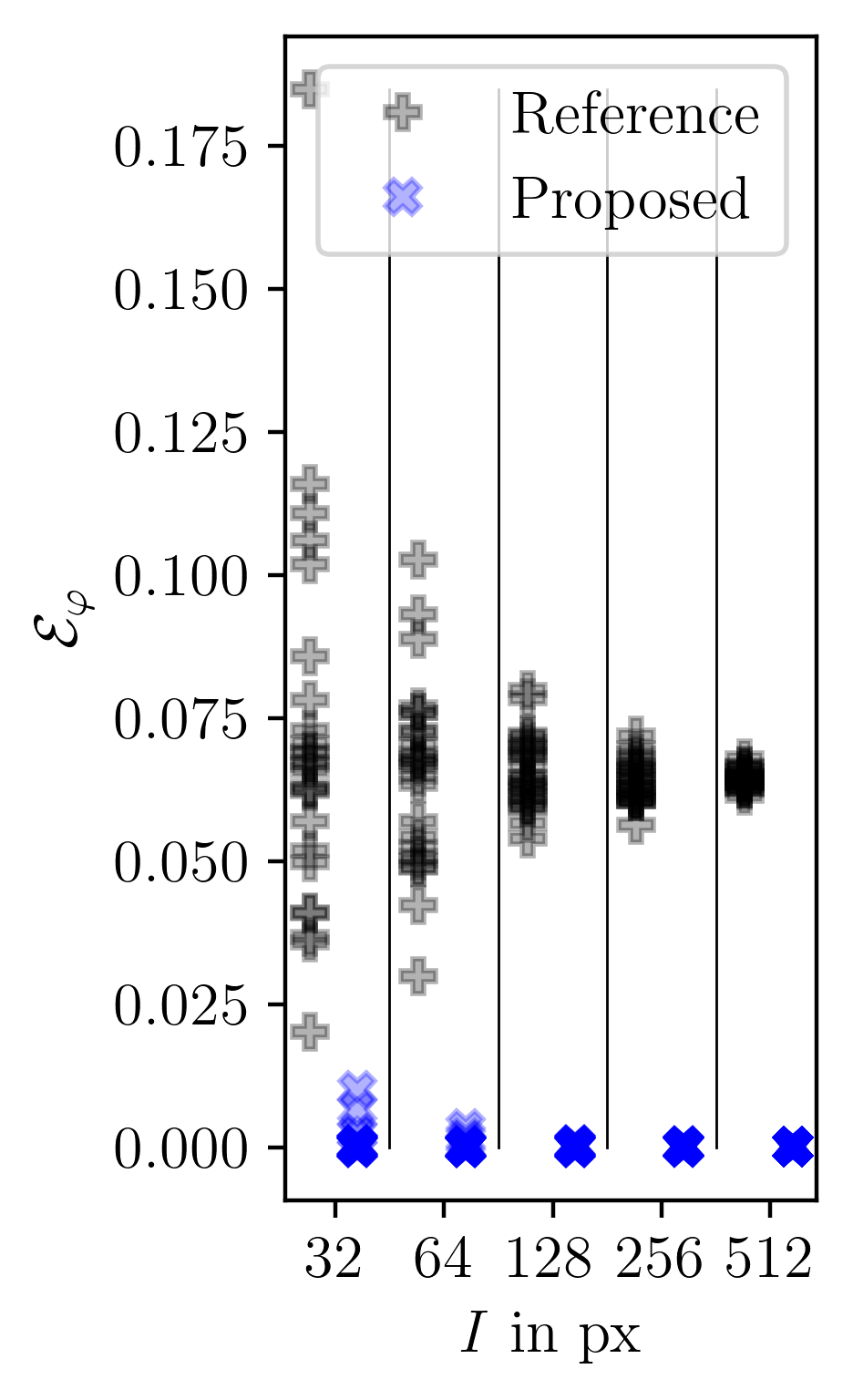}
\includegraphics[width=0.16\textwidth]{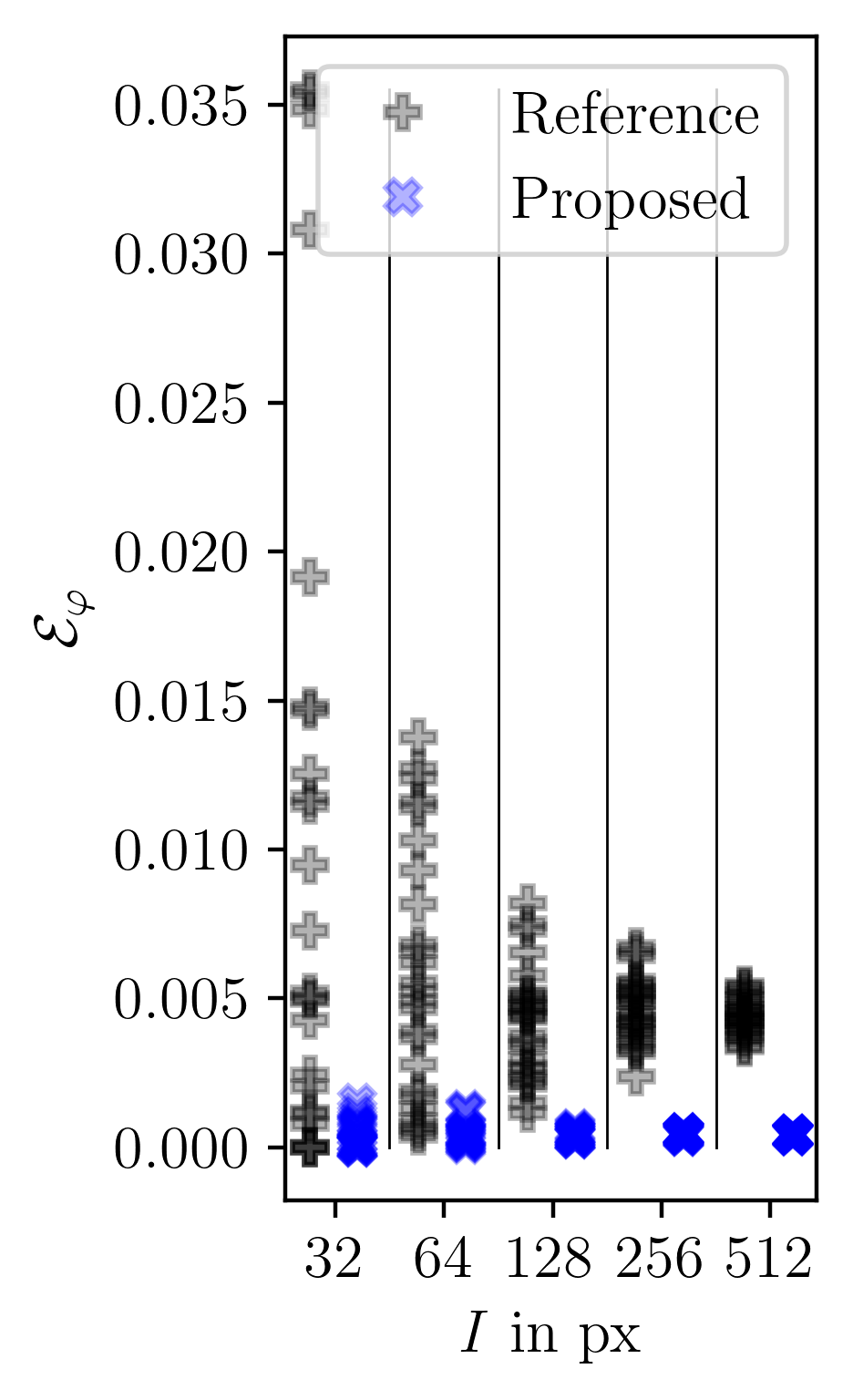}
\\
\includegraphics[width=0.16\textwidth]{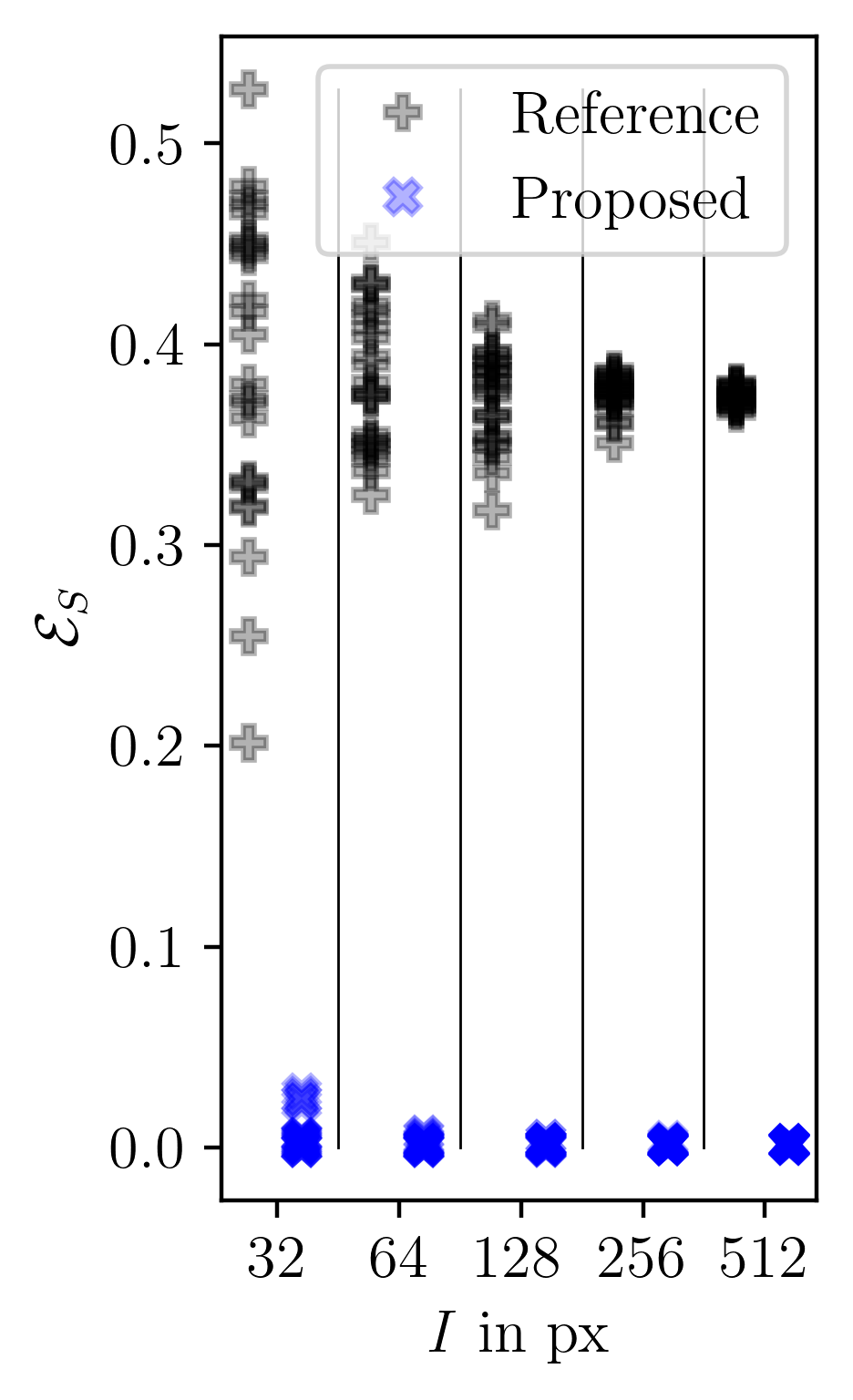}
\includegraphics[width=0.16\textwidth]{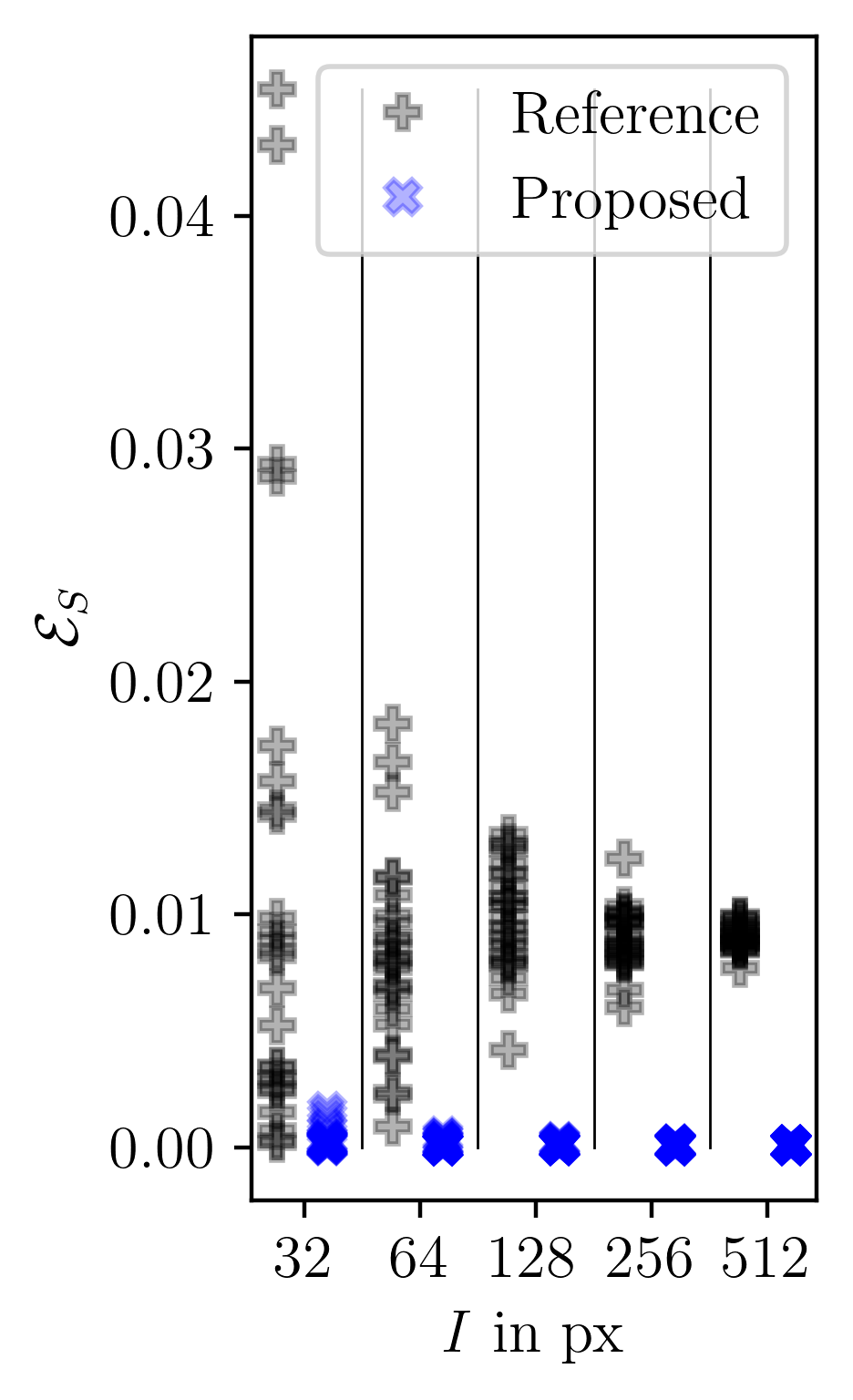}
\includegraphics[width=0.16\textwidth]{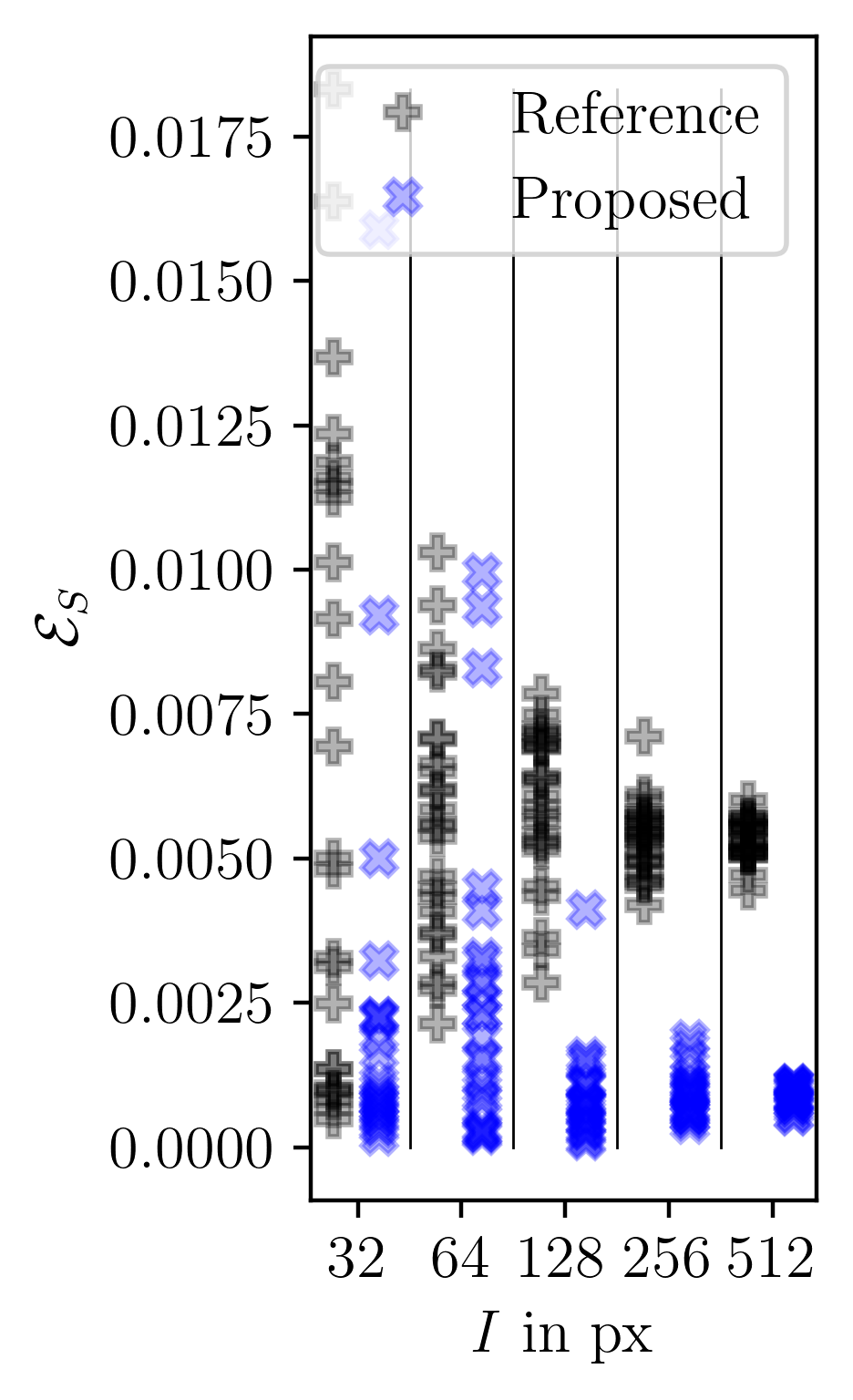}
\includegraphics[width=0.16\textwidth]{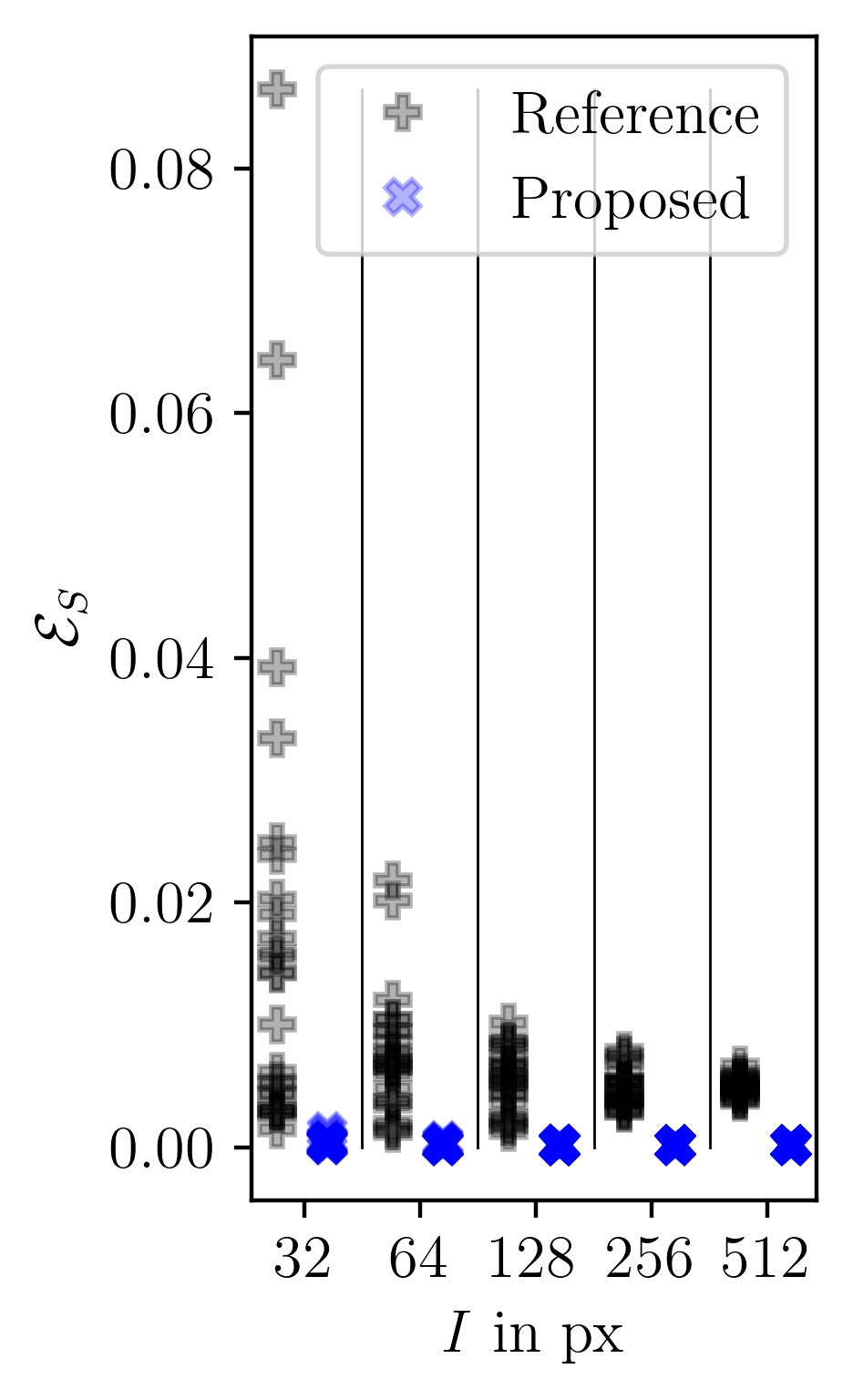}
\includegraphics[width=0.16\textwidth]{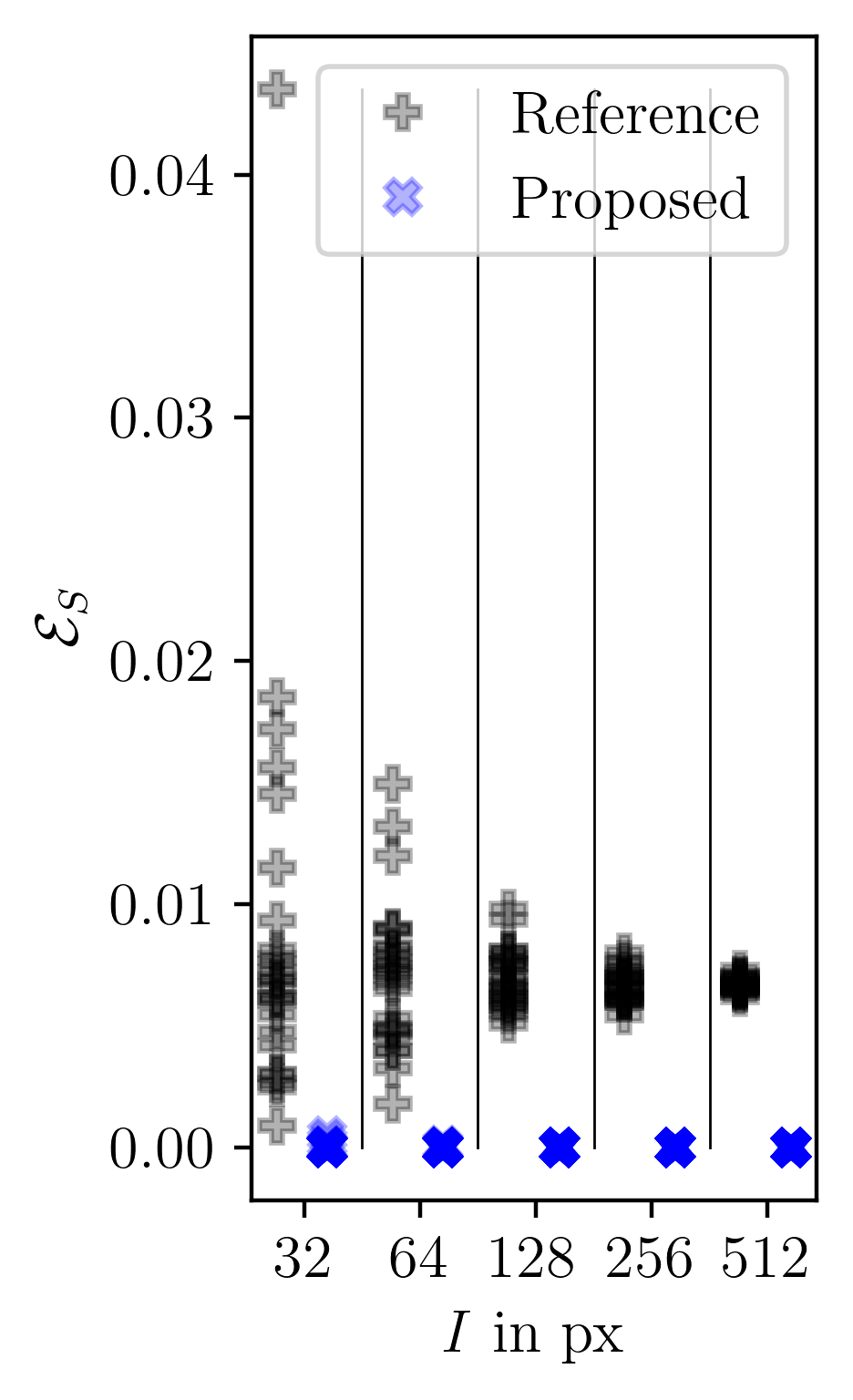}
\includegraphics[width=0.16\textwidth]{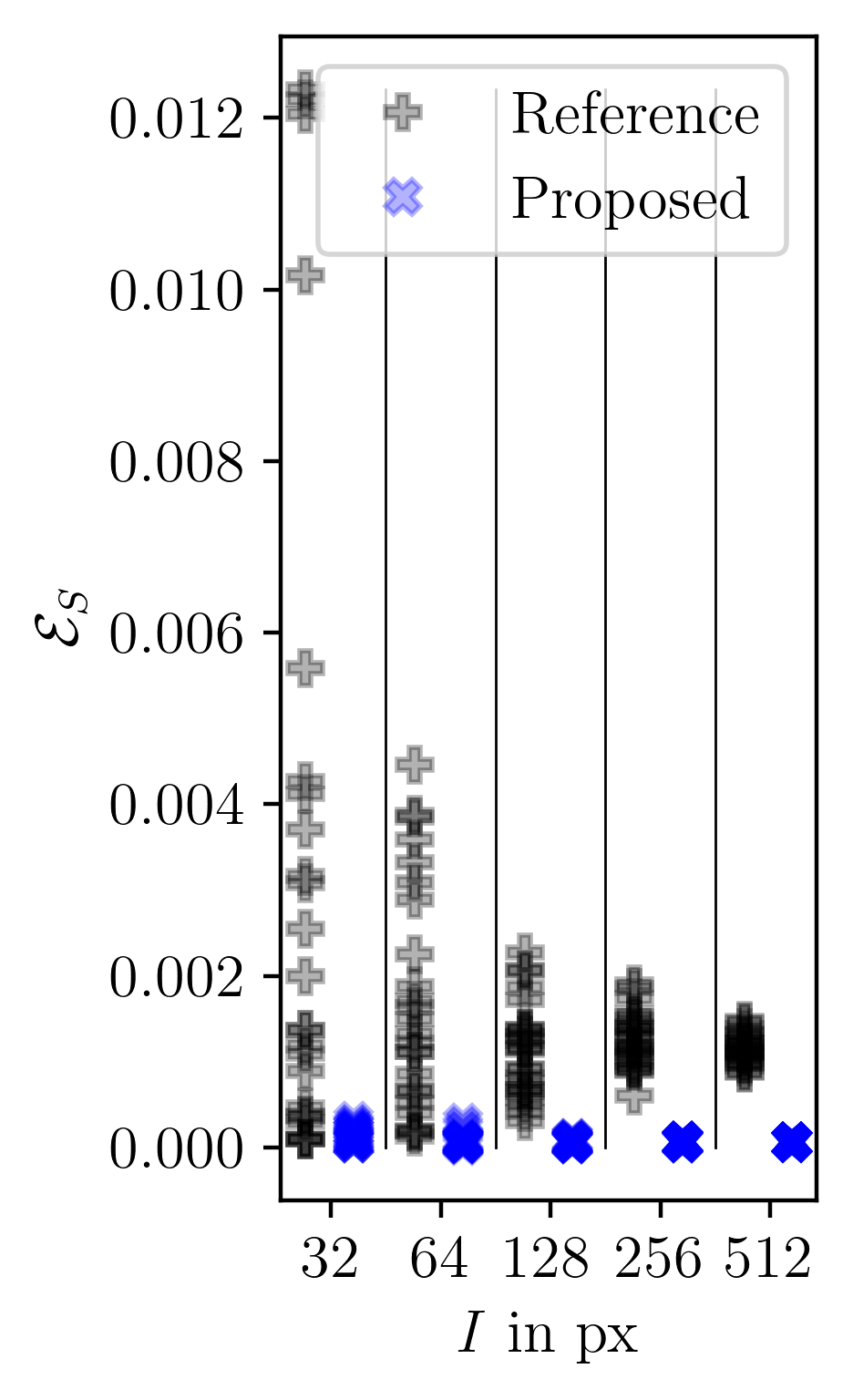}
\\
\subfigure[Alloy]{\includegraphics[width=0.16\textwidth]{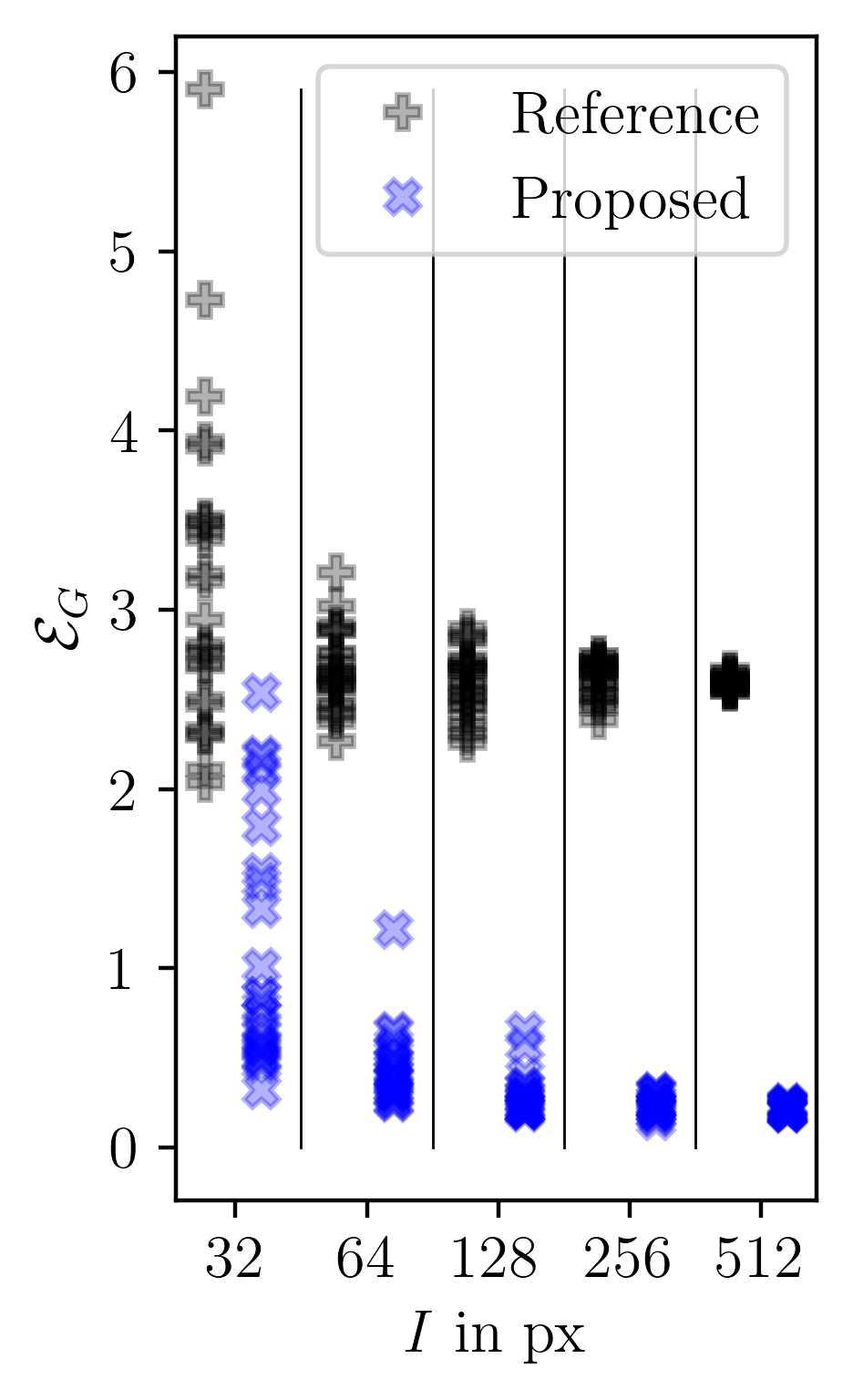}}
\subfigure[Carbonate]{\includegraphics[width=0.16\textwidth]{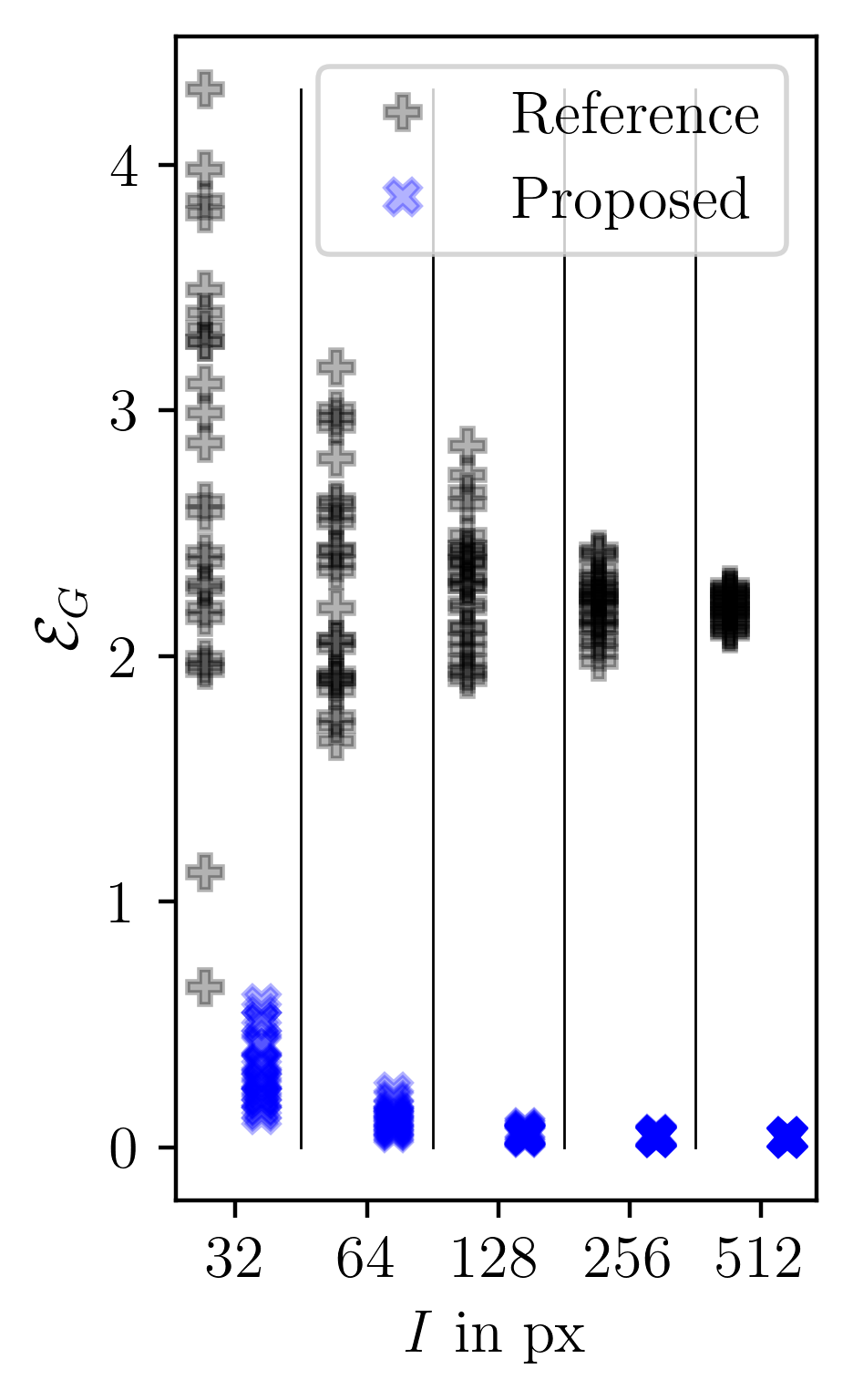}}
\subfigure[Ceramics]{\includegraphics[width=0.16\textwidth]{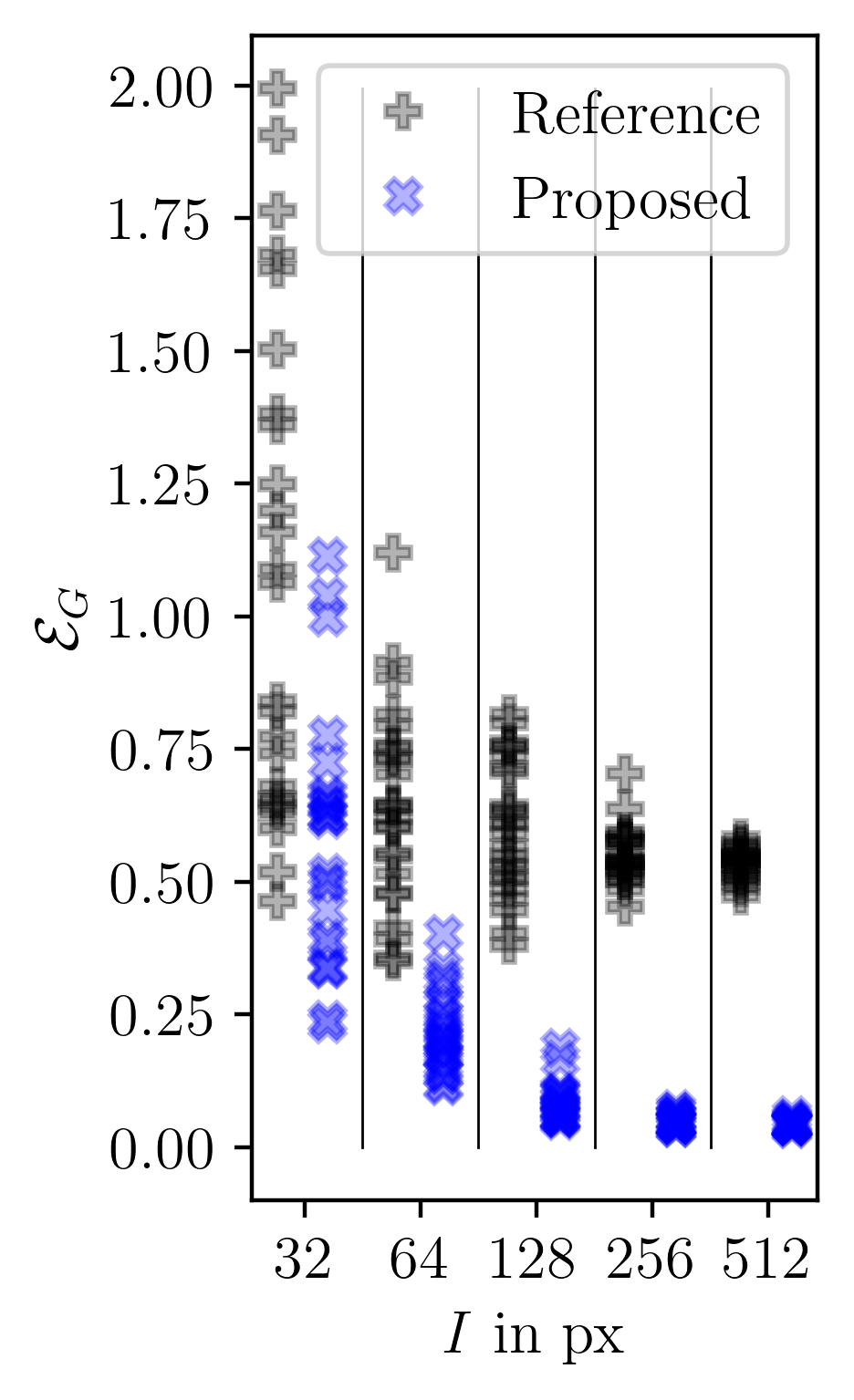}}
\subfigure[Copolymer]{\includegraphics[width=0.16\textwidth]{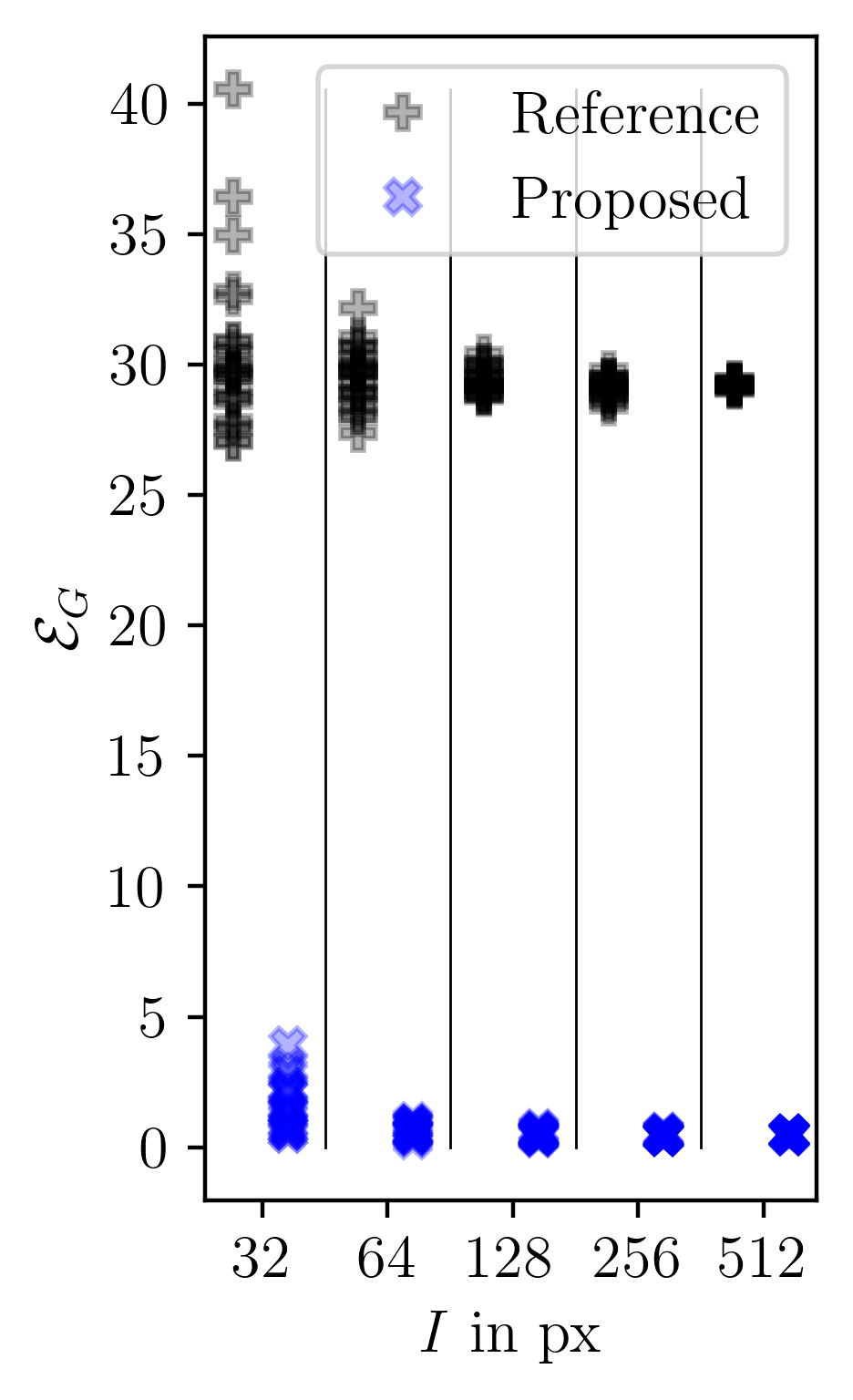}}
\subfigure[PMMA]{\includegraphics[width=0.16\textwidth]{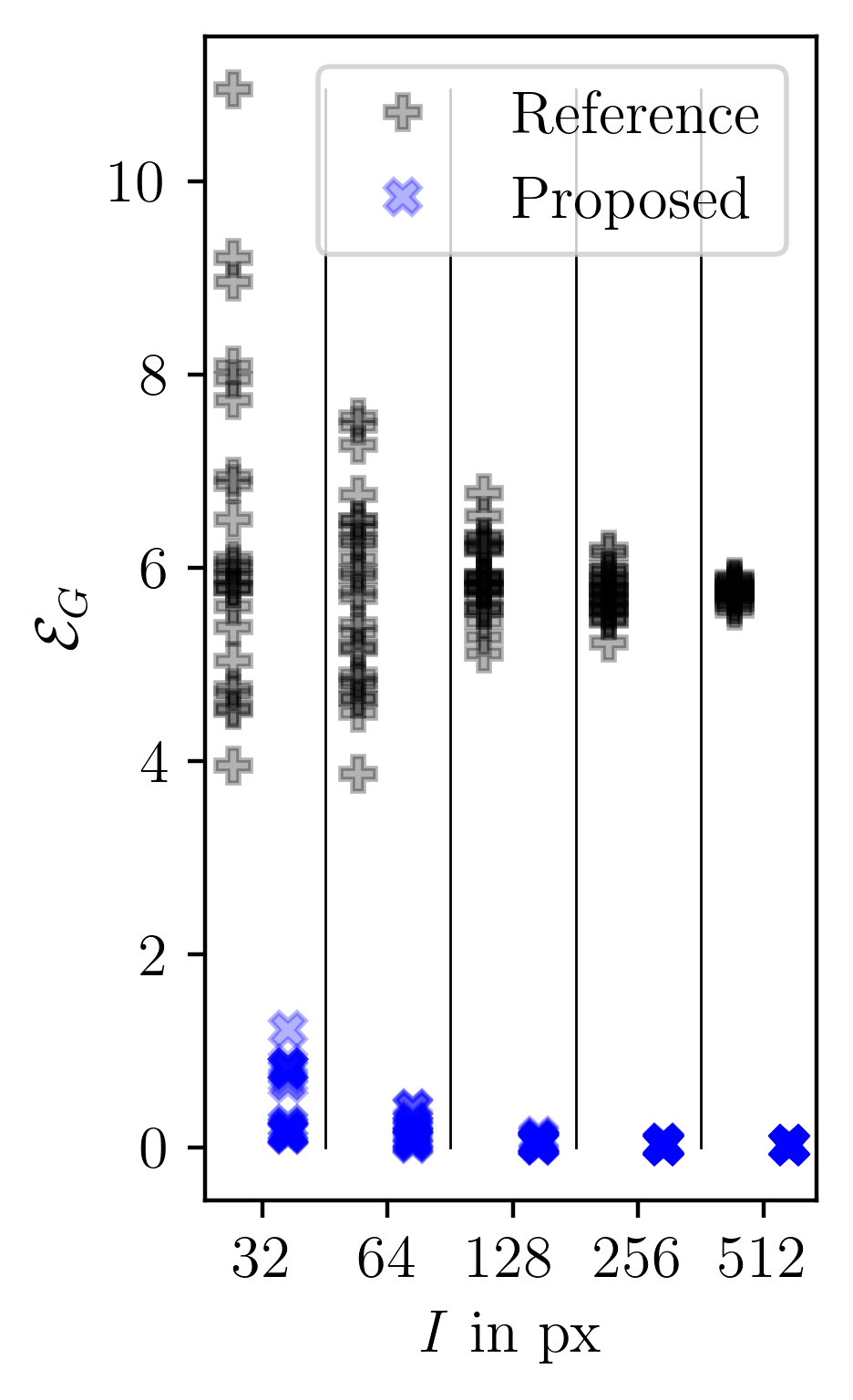}}
\subfigure[Sandstone]{\includegraphics[width=0.16\textwidth]{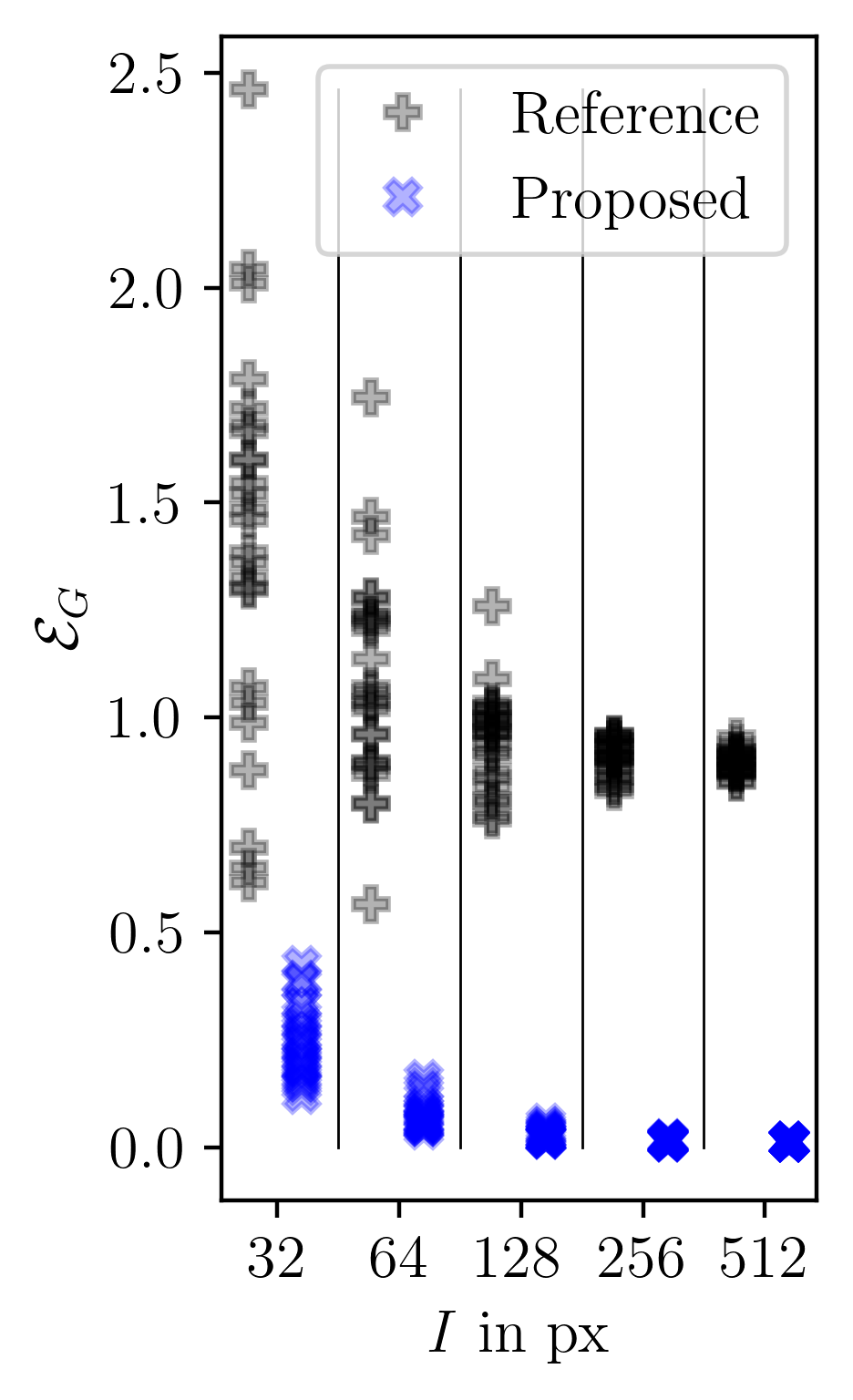}}
\caption{Influence of the loss function on the descriptor errors. The volume fractions~$\varphi$ (top), spatial correlations~$S$ (middle) and Gram matrices~$G$ (bottom) are compared for different materials (left to right) with 25 realizations per resolution. The resolutions are powers of two and an offset to the left (reference) and right (proposed) is applied only for visualization purposes. If can be seen that regardless of the model, material and descriptor, the variance of the descriptor error over different realizations decreases as the sample size increases. The proposed model consistently outperforms the reference~\cite{mordvintsev_texture_2021}. For some structures like the ally (a), the reference model fails to converge, leading to massive discrepancies, whereas for the ceramics (c) the differences are relatively small.}\label{fig:descriptor_convergence}
\end{figure*}

This is shown in \autoref{fig:descriptor_convergence}, where the error
\begin{equation}
    \mathcal{E}_D = \| \boldsymbol{D}(m^\mathrm{end}) - \boldsymbol{D}^\mathrm{des} \|_\mathrm{MSE}
\end{equation}
between a descriptor~$\boldsymbol{D}(m^\mathrm{end})$ and its desired value~$\boldsymbol{D}^\mathrm{des}$ from the reference structure is defined as a mean squared error (MSE).
Note that the only difference between~$\mathcal{E}_D$ and the loss~$\mathcal{L}$ defined in \autoref{eqn:loss} is that the former measures individual descriptors, whereas the latter is based on a weighted concatenation of multiple descriptors.
For all tested descriptors, the error converges to a value which is consistently lower for the proposed loss model than for the reference NCA-based texture synthesis method by Mordvintsev et al.~\cite{mordvintsev_texture_2021}.
It should be noted that the descriptor errors do not converge to zero as the resolution increases, but rather to a value that depends on the training quality.
It is observed that longer training and training by larger samples reduces this value (not shown here).

\begin{figure*}[htpb]%
\centering
\includegraphics[width=\textwidth]{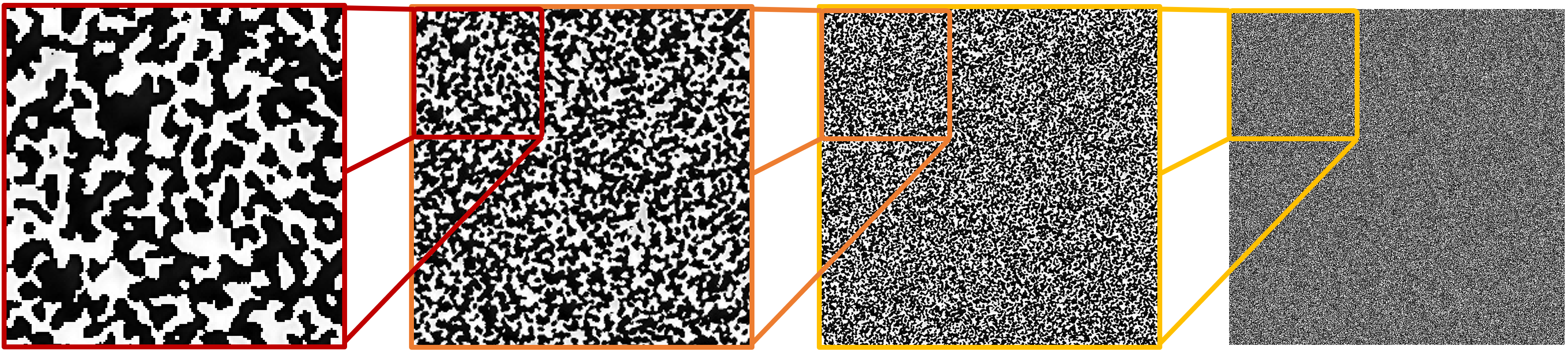}
\caption{A reconstruction of the ceramics microstructure (Figure \autoref{fig:various_reconstructions:ceramics}) with $4096^2$ pixels shows the scalability of the method. Three different zoom levels of the same structure are shown for visualization purposes.}\label{fig:large}
\end{figure*}

An interesting aspect of NCA is that the image sizes during training and sampling are independent. 
This is favorable because the sampling is relatively inexpensive and scales favorably with the image size compared to other methods.
To demonstrate this, \autoref{fig:large} shows a reconstruction example where the resolution is chosen such that the sampling takes as long as the training\footnote{Because the utilized VRAM is not sufficient for the large reconstruction, it is conducted on the CPU only, whereas the training occurs on the GPU. If a similar comparison was made on identical hardware, significantly larger structures could be reconstructed. Regardless, in the authors' opinion, the presented results demonstrate the scalability sufficiently well.}.
Three different zoom levels of the same structure are shown for visualization purposes.
Without any multigrid procedures, such large reconstructions are very challenging with classical descriptor-based methods.

\begin{figure}[htpb]%
\centering
\includegraphics[width=0.85\linewidth]{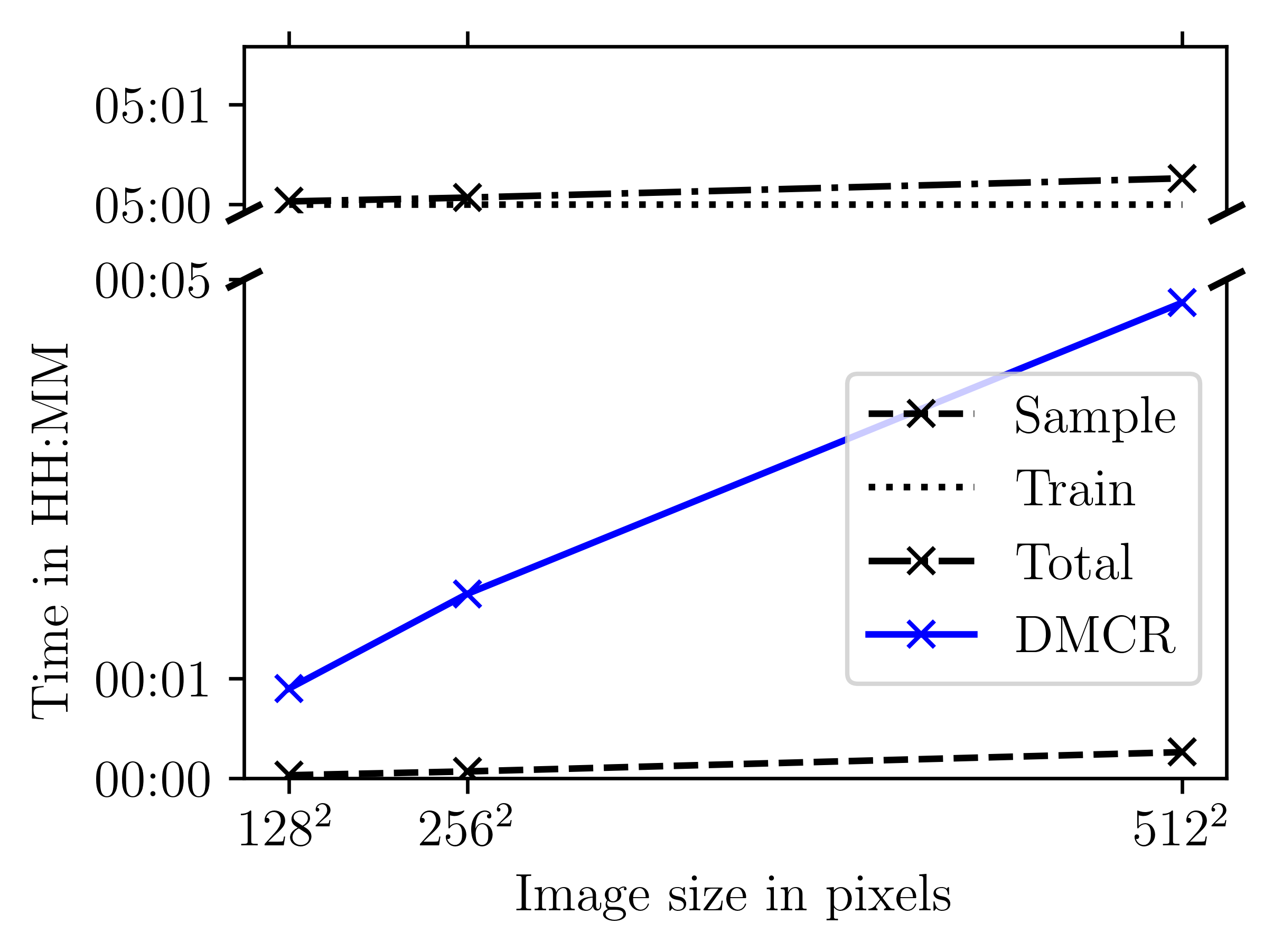}
\caption{Scalability comparison of descriptor-based NCA with the 2D DMCR algorithm~\cite{seibert_reconstructing_2021} in \emph{MCRpy}~\cite{seibert_microstructure_2022}. Both scale linearly in the number of pixels, which scales quadratically in the side length. The constant offset of the NCA cost due to training is compensated by the much faster sampling, which also grows at a slower rate.}\label{fig:scalability}
\end{figure}

Generally, the computational cost of sampling a microstructure scales linearly in the number of pixels, because pixel updates are computed independently by the NCA. 
A comparison with the 2D DMCR algorithm~\cite{seibert_reconstructing_2021} in \emph{MCRpy}~\cite{seibert_microstructure_2022} is given in \autoref{fig:scalability}.
Both methods scale linearly.
Sampling from a trained NCA is much faster than reconstructing by DMCR, and the computational cost grows more slowly.
This is because the expensive evaluation of microstructure descriptors and iterative optimization are moved to the training stage.
If the training is added to the computational cost of the NCA, they are slower for the considered microstructure sizes.
As a conclusion, the expensive training phase of an NCA is compensated if large or many microstructures are reconstructed.
Especially the latter might speed up a potential future extension for 2D-to-3D reconstruction.
Furthermore, unlike with DMCR~\cite{seibert_reconstructing_2021,seibert_descriptor-based_2022}, the sampling can be trivially parallelized because updates are based only on local information.

\section{Conclusion}\label{sec:conclusion}
A neural cellular automaton (NCA)-based algorithm for microstructure reconstruction is presented.
The microstructure evolution is modeled as a partial differential equation which is learned by a small neural network, the NCA.
Despite the purely local information in the NCA, long-range correlations are incorporated by introducing hidden dimensions to the microstructure function which can be used to communicate information.
Unlike with previous approaches, this network is not trained on image data but on statistical microstructure descriptors.
Thus, the method incorporates ideas from four different families of microstructure generation approaches, namely simulation, Markov, deep learning and descriptor-based methods, which are all briefly reviewed.
The method is formulated, implemented and validated by a number of 2D numerical experiments.

Compared to other microstructure reconstruction approaches, descriptor-based NCAs have a unique set of advantages.
The neural network in the NCA enables the evolution of highly complex morphologies in a PDE-like manner without knowledge of the governing physical equations and the material parameters.
It can be controlled by statistical descriptors.
However, the sampling of structures from a trained NCA is based only on local information.
This self-assembling nature of the algorithm makes it an inherently distributed algorithm and therefore trivial to parallelize.
The random selection of the pixels to be updated make the method robust with respect to random perturbations, as long as not all channels are affected.
Finally, the method scales very favorably as arbitrarily resolved structures can be sampled.

In future work, the main challenge lies in enabling 3D reconstruction based on 2D or 3D reference data.

\section*{Acknowledgements}
The authors thank Anastasia Opara for providing good implementations of texture synthesis algorithms to the community.
The groups of M. Kästner and D. Peterseim thank the German Research Foundation DFG which supported this work under Grant numbers KA 3309/18-1 and PE 2143/7-1, respectively.



\section*{Code and data availability}
The code is made available upon reasonable request.

The data is taken from the literature~\cite{li_transfer_2018}, where it is released under the Creative Commons license.

\section*{Competing interests}
The authors declare no competing interests.

\section*{Author contributions}
\textbf{P. Seibert}: Conceptualization, Data Curation, Formal Analysis, Investigation, Methodology, Software, Supervision, Validation, Visualization, Writing - Original Draft Preparation, Writing - Review \& Editing.
\textbf{A. Raßloff}: Conceptualization, Writing - Review \& Editing.
\textbf{Y. Zhang}: Software, Writing - Review \& Editing.
\textbf{K. Kalina}: Conceptualization, Formal Analysis, Writing - Review \& Editing.
\textbf{P. Reck}: Conceptualization, Writing - Review \& Editing.
\textbf{D. Peterseim}: Conceptualization, Funding Acquisition, Writing - Review \& Editing.
\textbf{M. Kästner}: Conceptualization, Funding Acquisition, Resources, Supervision, Writing - Review \& Editing.


\bibliographystyle{spphys}
{\small \bibliography{preprint}}

\end{document}